\documentstyle{mn}
\title{Time delay for the gravitational lens system B0218+357}
\author[A.~D.~Biggs et al.]{A.~D.~Biggs,$^1$ I.~W.~A. Browne,$^1$ 
P.~Helbig,$^1$ L.~V.~E.~Koopmans,$^2$ 
\newauthor P.~N.~Wilkinson,$^1$ and R.~A.~Perley$^3$\\
$^1$University of Manchester, Nuffield Radio Astronomy Laboratories,
Jodrell Bank, Macclesfield, Cheshire SK11 9DL, UK\\
$^2$University of Groningen, Kapteyn Astronomical Institute, Postbus 800, 
9700 AV Groningen, The Netherlands\\
$^3$National Radio Astronomy Observatory, P.O. Box 0, Socorro, NM 87801, USA}
\begin{document}
\maketitle
\begin{abstract}

Measurement of the time delay between multiple images of a gravitational
lens system is potentially an accurate method of determining the Hubble 
constant over cosmological distances. One of the most promising candidates 
for an application of this technique is the system B0218+357 which was 
found in the Jodrell Bank/VLA Astrometric Survey (JVAS). This system 
consists of two images of a compact radio source, separated by 335 
milliarcsec, and an Einstein ring which can provide a strong constraint on  
the mass distribution in the lens. We present here the results of a 
three-month VLA monitoring campaign at two frequencies. The data are 
of high quality and both images show clear variations in total flux 
density, percentage polarization and polarization position angle at 
both frequencies. The time delay between the variations in the two images 
has been calculated using a chi-squared minimization to be $10.5\pm0.4$ 
days at 95 per cent confidence, with the error being derived from 
Monte-Carlo simulations of the light curves. Although mass modelling of 
the system is at a preliminary stage, taking the lensing galaxy to be a 
singular isothermal ellipsoid and using the new value for the time delay gives 
a value for the Hubble constant of 69$^{+13}_{-19}$\,km\,s$^{-1}\,$Mpc$^{-1}$,
again at 95 per cent confidence.

\end{abstract}


\section{Introduction}

Long before the first gravitational lens was discovered, it had been 
shown \cite{refsdal64} that measurement of a time delay between the 
images of a lens could be used to calculate the Hubble constant, $H_0$, 
independently of any other distance determination to the lens or lensed 
object. This technique has so far been applied predominantly to the Double 
Quasar B0957+561 (Walsh, Carswell \& Weymann 1979) where a long-running 
controversy as to the length of the time delay has only recently been 
resolved \cite{kundic97,haarsma98}. In the first of these two papers, an 
$H_0$ of 64$\pm$13\,km\,s$^{-1}\,$Mpc$^{-1}$ was derived based on an optically 
determined time delay of 417$\pm$3 days (at 95 per cent confidence). However, 
the deflecting mass is complicated (comprising a galaxy and a galaxy cluster) 
and modelling it satisfactorily has proved difficult and constitutes the 
biggest source of error on the value of $H_0$ derived from this system at 
this time. 

\begin{figure}
\begin{center}
\setlength{\unitlength}{1cm}
\begin{picture}(5,6)(0,0)
\put(-1.3,-2){\includegraphics{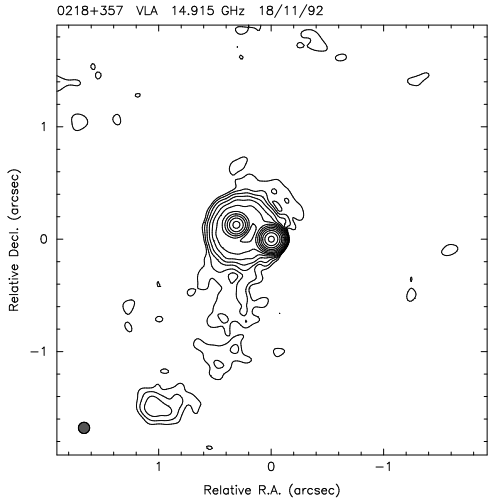}}
\end{picture}
\caption{VLA 15~GHz radio map of B0218+357. As well as the two compact 
components (A to the right) and the Einstein ring, also clearly visible 
is a (non-lensed) radio jet to the south.}
\label{0218map}
\end{center}
\end{figure}

Systems containing an Einstein ring are ideal candidates for determining 
$H_0$ as the presence of the ring can firmly constrain the mass 
distribution in the lens \cite{kochanek90}. A good example of this 
is B0218+357 \cite{patnaik93}, first identified as a gravitational lens 
through observations carried out as part of the Jodrell Bank/VLA Astrometric 
Survey (JVAS) \cite{patnaik92}. This lens system has a 
simple morphology (see Fig.~\ref{0218map}) which consists of two 
compact images (A and B) of a strongly variable flat-spectrum radio 
core and a steep-spectrum Einstein ring, the diameter of which is the 
same as the separation of the compact components, 335 milliarcsec (mas). 
This is the smallest separation yet found in a galactic-mass gravitational 
lens system and as a consequence, the time delay between the variations 
in components A and B is small. The ring is believed to be an image of 
part of the extended structure of the kpc-scale radio jet and will 
therefore vary on much 
longer time scales than the variations seen in the images of the compact 
cores. The deflecting mass comprises a single isolated galaxy which, in 
contrast to B0957+561, can be modelled relatively simply. The galaxy is 
also almost certainly a spiral because radio absorption observations 
have shown that the column density of absorbing material is very 
high (Carilli, Rupen \& Yanny 1993, Wiklind \& Combes 1995) and because 
a large differential Faraday rotation measure exists between A and B at 
radio wavelengths \cite{patnaik93}. The redshifts of the lensed object and 
lensing galaxy are well determined at 0.96 \cite{lawrence96} and 
0.6847 \cite{ian93} respectively. As these are relatively low for a 
lens system, the assumed cosmology introduces less uncertainty in 
the value of $H_0$ than with other systems.

Previous work \cite{liz96} derived a time delay of 12 days (B lagging A) 
with a 1$\sigma$ error of $\pm$3 days from VLA observations of the 
percentage polarized flux at 15~GHz. In this paper we present new 
results of a three-month monitoring campaign conducted with the VLA at 
8.4 and 15~GHz.

\section{Observations and Data Reduction} 

B0218+357 was observed with the VLA in A configuration between the 
months of October 1996 and January 1997. Observations were taken at 
two frequencies, 15~GHz and 8.4~GHz, each with a bandwidth of 50 MHz. 
Each band is further split into two {\em intermediate frequencies}, or 
IFs, separated slightly in frequency. With resolutions of 120 mas at 
15~GHz and 200 mas at 8.4~GHz, the variable components A and B are 
easily resolved and can be monitored for variations in total flux density, 
percentage linear polarization and polarization position angle. In all, 
data were obtained at 47 epochs, with an average spacing between 
observations of $\sim$2 days. 

The observing strategy was kept as consistent as possible over the 
period of the monitoring. Each epoch contained two observations of 3C84,
one before and one after B0218+357, for amplitude and phase calibration 
purposes. The 3C84 data are also used to correct for instrumental 
polarization on the assumption that 3C84 is unpolarized. 3C119 was observed 
as a `control source' to check for variations in the flux density of 
3C84 (each epoch is calibrated assuming the same flux for 3C84) and to 
calibrate the polarization position angle of
B0218+357. 3C119 is a steep-spectrum source known to contain a very weak 
core \cite{rendong91} and so any variations in its total flux density 
should be very small. An innovation compared to the previous monitoring 
campaign was that antenna pointing offset observations were made of each 
source to ensure that gain variations associated with pointing errors 
were minimized. B0218+357 was observed for $\sim$20 minutes at 15~GHz 
and $\sim$3 minutes at 8.4~GHz resulting in expected rms noise levels 
of 0.19 mJy and 0.12 mJy respectively for one IF.

Calibration was performed using the NRAO Astronomical Image Processing 
Software package {\sc aips} and each IF was calibrated separately. As 
the primary amplitude and phase calibrator, 3C84, is slightly resolved 
at the observed frequencies, the first step in the calibration process 
was to make the best possible map of 3C84. This was done by performing 
a first order calibration of several epochs of 3C84 data and combining these 
to produce a self-calibrated map of the source. The clean components so 
derived were then used as a model input to the {\sc aips} task {\sc calib} 
to derive telescope gain solutions at each epoch of observation. These 
gain solutions were then applied to each of the target sources, 3C119 
and B0218+357. The calibration of the 15~GHz total flux density data 
was improved significantly by correcting for the gain of the VLA antennas 
changing with elevation. At lower frequencies the gain/elevation 
dependence is much weaker and at 8.4~GHz a correction made little 
or no difference to the gain solutions. No correction at this frequency 
has therefore been applied.

The flux densities of components A and B of B0218+357 were calculated by 
fitting a model of the source directly to the calibrated {\em uv} 
visibility data using the {\sc difmap} \cite{shepherd97} model fitter 
(for Stokes $I$) and its {\sc aips} equivalent {\sc uvfit} 
(Stokes $Q$ and $U$). The model was kept as simple as possible and 
consisted of two point sources (separated by 335 mas at a position 
angle of 67 degrees) plus a broad Gaussian to represent the ring. 
The model fitting consisted of varying only the flux densities of 
these three components. Whereas the 
positions of components A and B are known to very high accuracy from 
VLBI observations, the size and position of the ring at each frequency 
was determined by performing an initial fit to each epoch's data and 
allowing the ring parameters to float. As no systematic variations in 
the ring parameters were seen, the size and position of the broad 
Gaussian were then fixed at the average value derived from all the epochs.

Visually the model provides a good fit to the data at all epochs, 
especially at 15~GHz as the ring flux density is low and the ring 
heavily resolved on most of the baselines. The $\chi^2$ per degree 
of freedom ($\overline{\chi}^2$) of fits at this frequency were 
typically less than 2. At 8.4~GHz the ring emission is stronger 
relative to components A and B and less separated from the point 
sources, resulting in poorer fits ($\overline{\chi}^2$ $\sim$ 3--7). 
In the case of 3C119 the source can be reasonably approximated by 
a single Gaussian. 

At this point we average the results found for the two IFs.
Bad epochs could be identified by a particularly high $\overline{\chi}^2$ 
or through inconsistent ring or control source flux densities and/or 
position angles. In most cases, the problem was then traced back to 
the original data and rectified. Because of either bad weather or 
instrumental problems, eleven 15~GHz and eight 8.4~GHz epochs of total 
flux density measurements have been removed from the time delay analysis. 
A particular problem occurred with the pointing offset observations for 
the first three epochs which has rendered the total flux density data 
unusable. However, the polarization information was obtainable for 
all but one of the epochs lost in total flux by amplitude 
self-calibrating the data, the results of which seem compatible with 
the other epochs. This gives 45 epochs of polarization data at 15~GHz 
and 44 at 8.4~GHz. 

\section{The Radio `Light Curves'}

\begin{figure*}
\begin{center}
\setlength{\unitlength}{1cm}
\begin{picture}(20,22)(0,0)
\put(1,21) {a)}
\put(9.5,14){\includegraphics{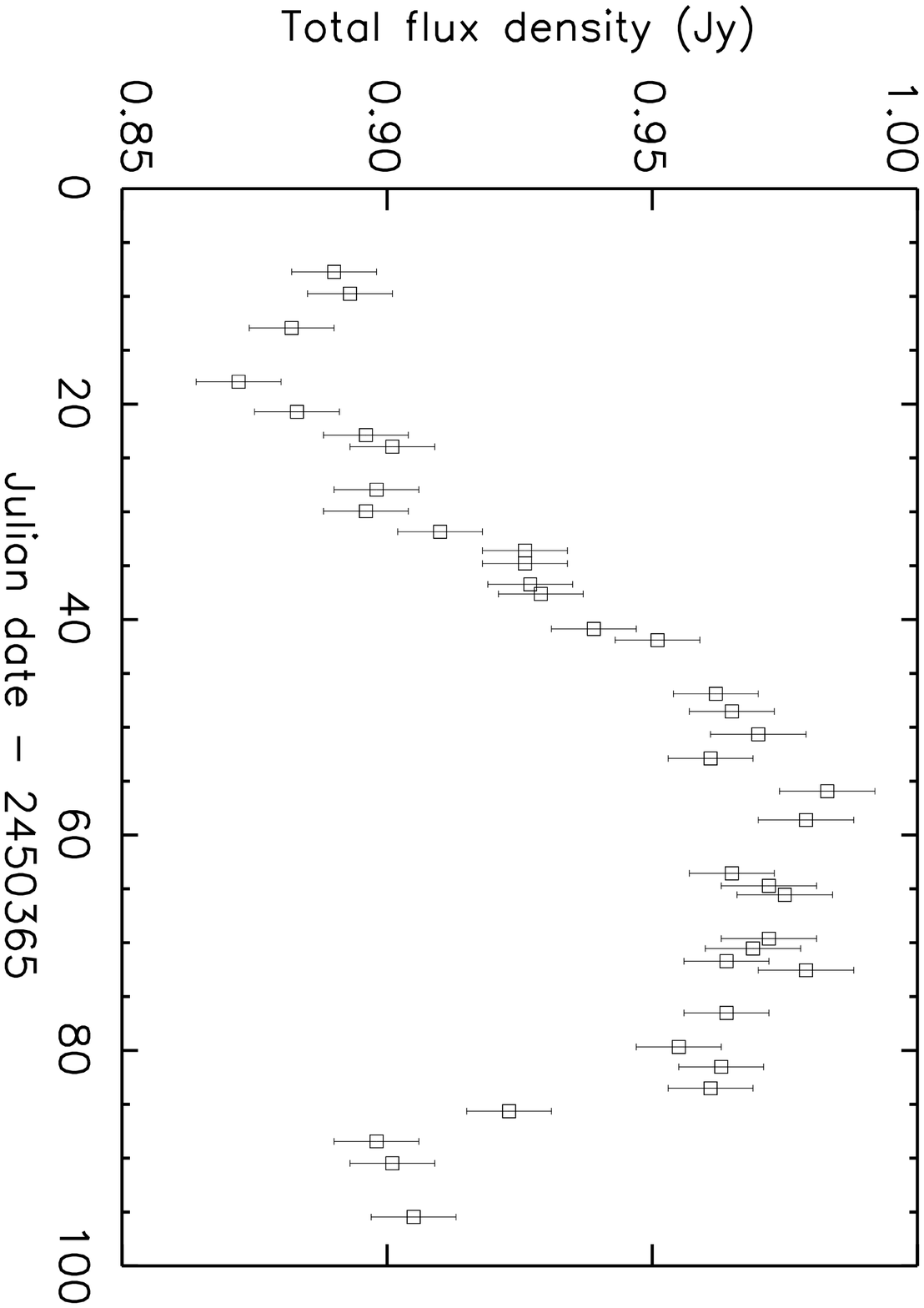}}
\put(10,21) {b)}
\put(18.5,14){\includegraphics{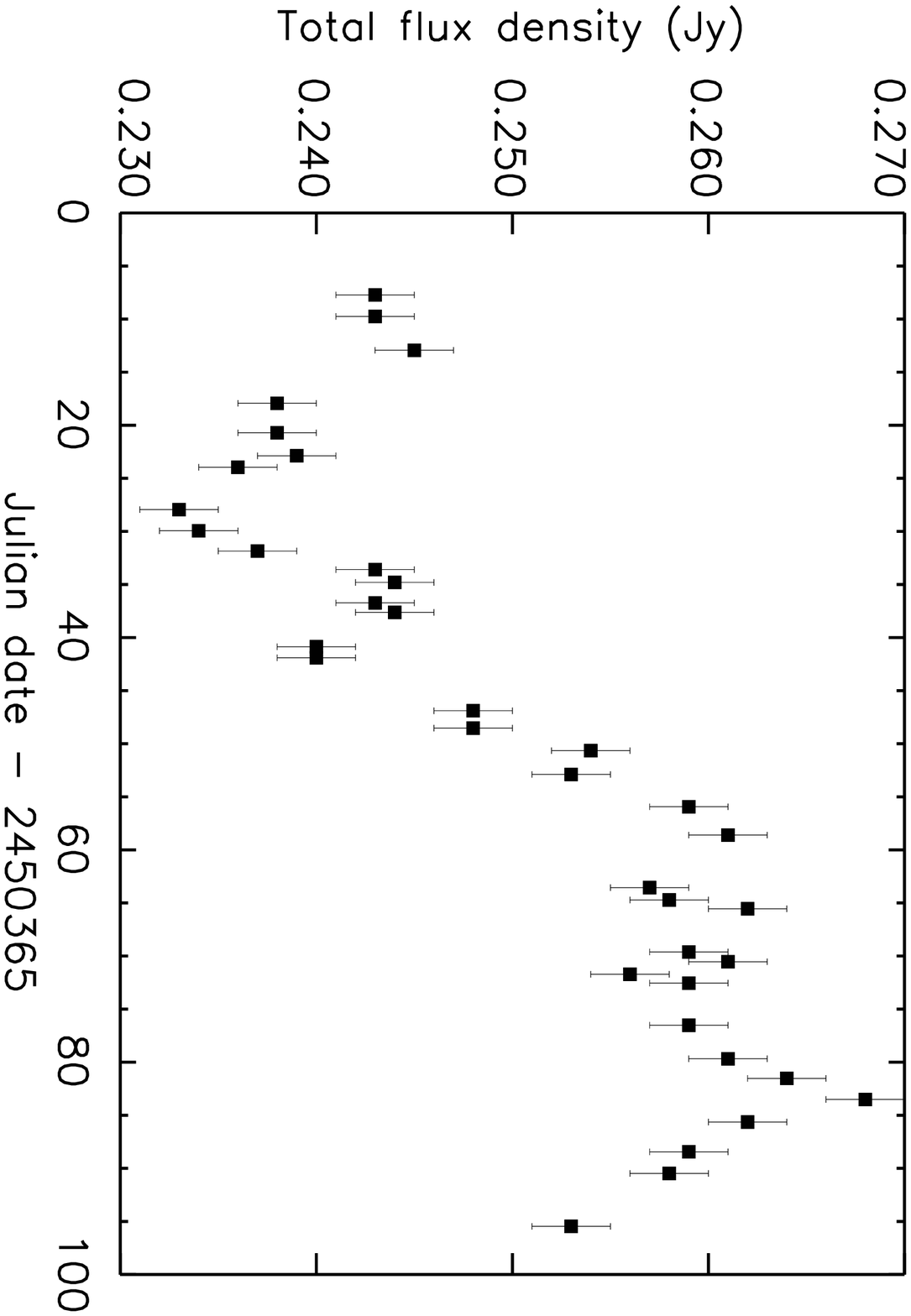}}
\put(1,14) {c)}
\put(9.5,7){\includegraphics{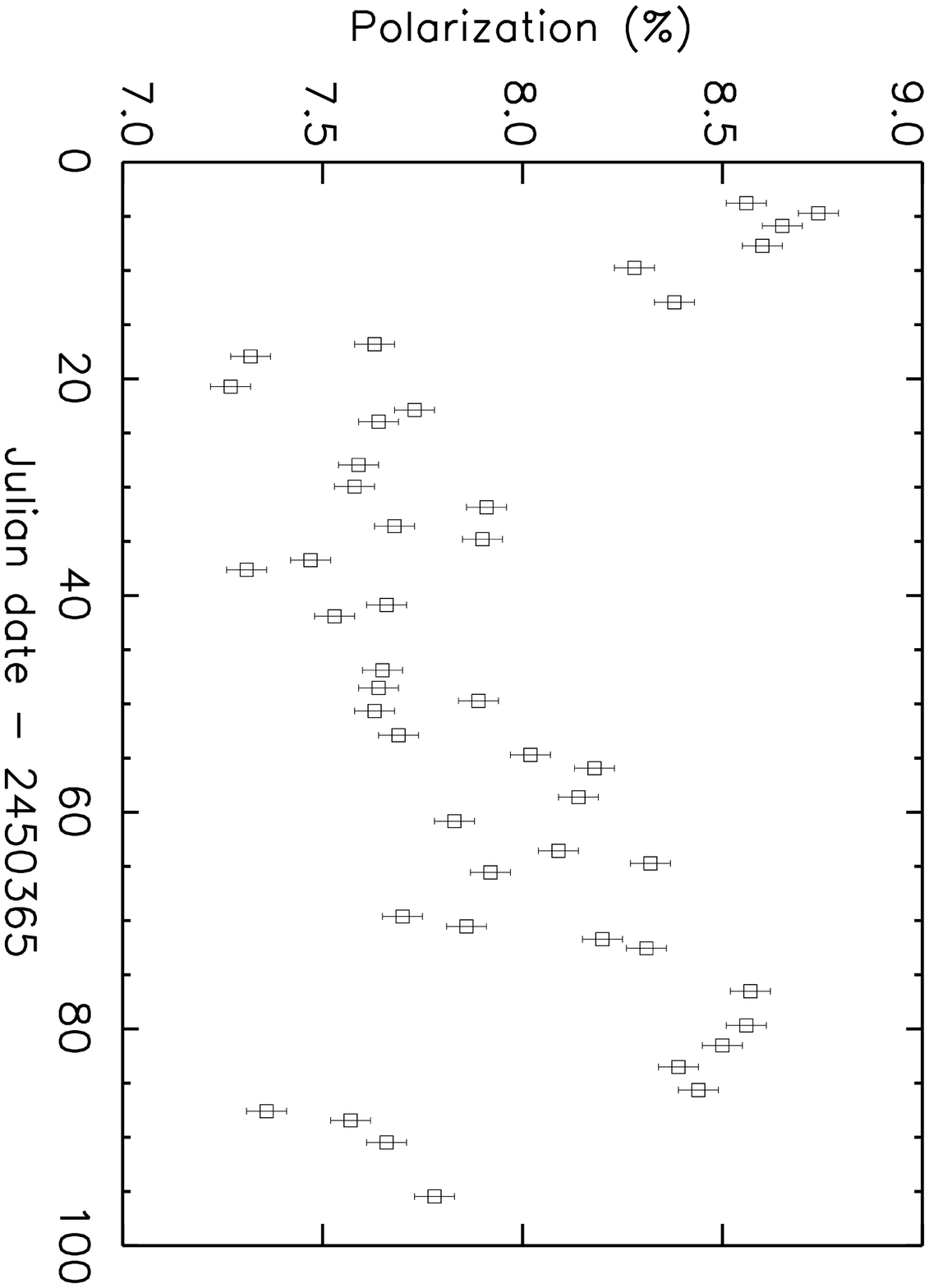}}
\put(10,14) {d)}
\put(18.5,7){\includegraphics{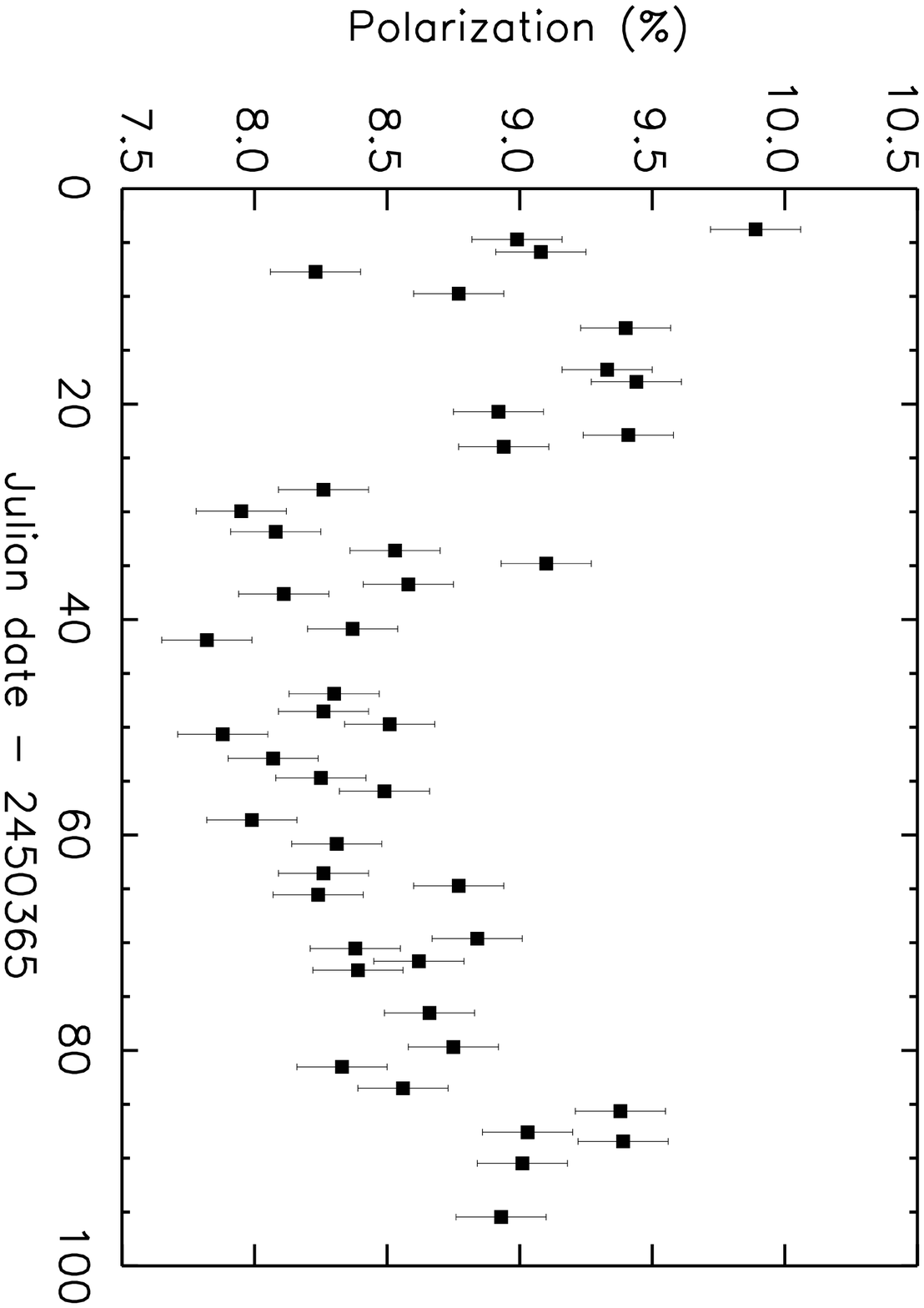}}
\put(1,7) {e)}
\put(9.5,0){\includegraphics{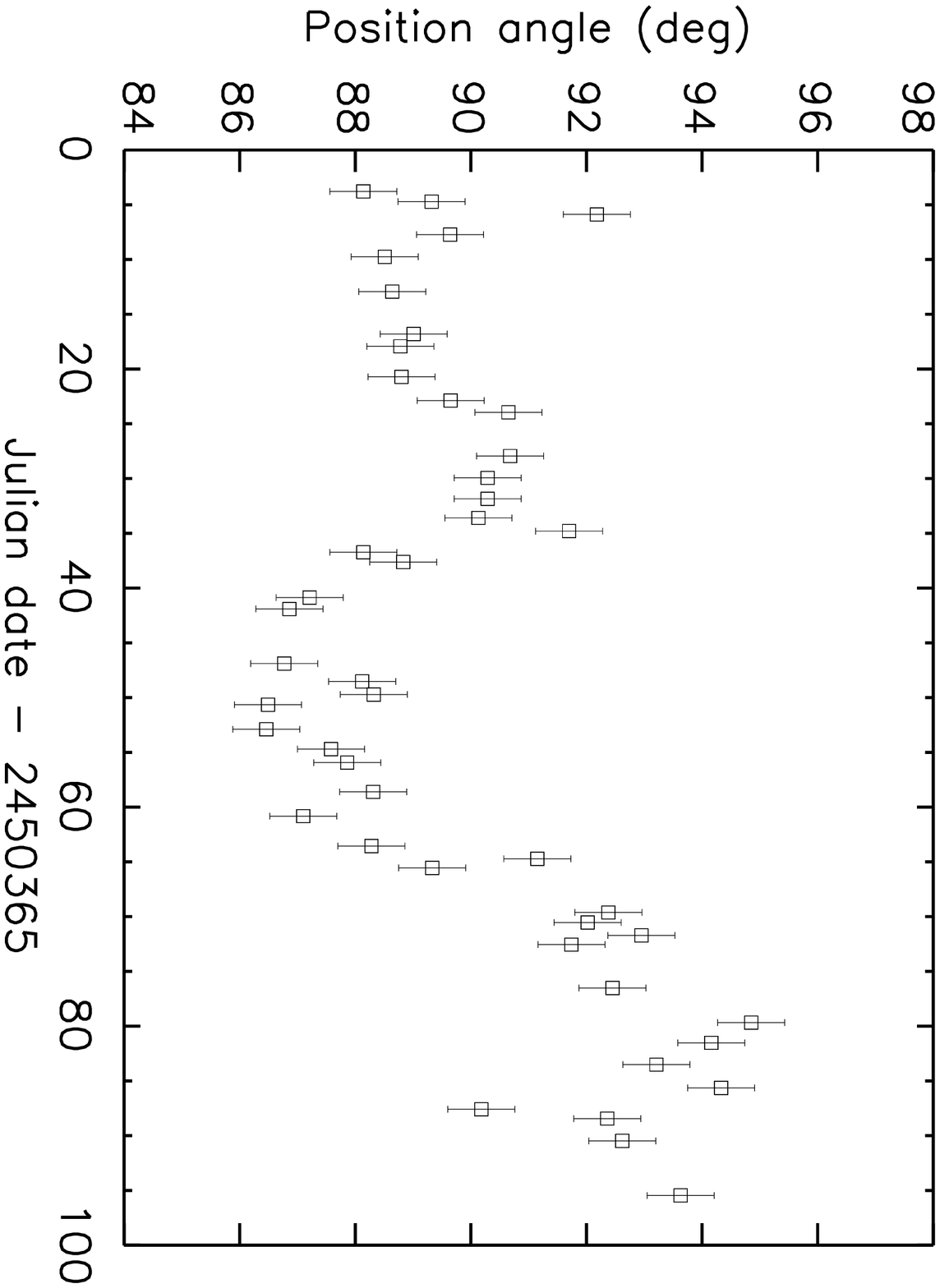}}
\put(10,7) {f)}
\put(18.5,0){\includegraphics{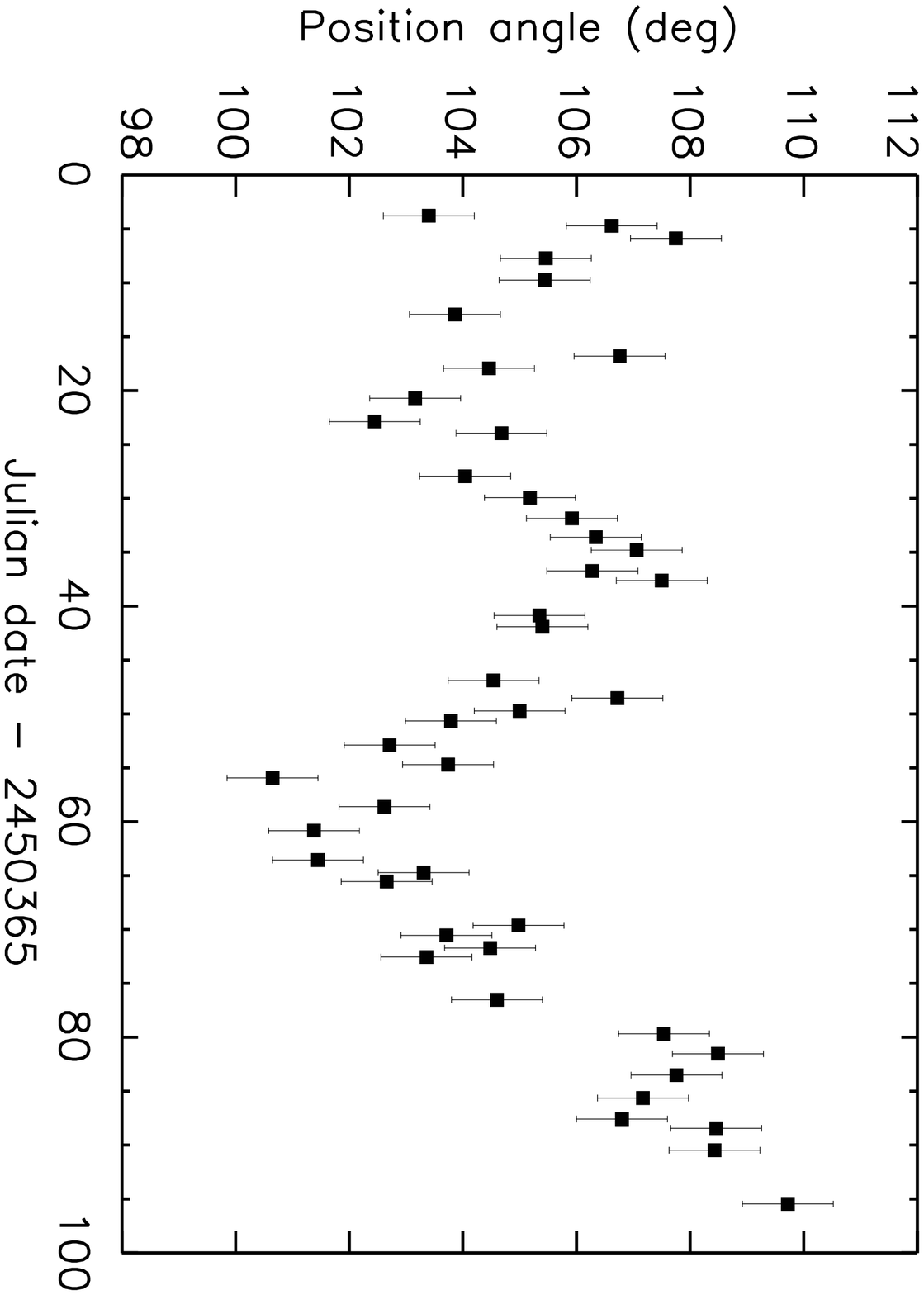}}
\end{picture}
\caption{15~GHz light curves. Top row: Total flux density of B0218+357, a) 
component A, b) component B. Middle row: Percentage polarization, 
c) component A, d) component B. Bottom row: Polarization position angle, 
e) component A, f) component B. (The data plotted in 
Figs~\ref{Ugraphs},~\ref{Xgraphs} and~\ref{3C119graphs} can be found at 
http://multivac.jb.man.ac.uk:8000/ceres/data\_from\_papers/0218.html)}
\label{Ugraphs}
\end{center}
\end{figure*}

\begin{figure*}
\begin{center}
\setlength{\unitlength}{1cm}
\begin{picture}(20,22)(0,0)
\put(1,21) {a)}
\put(9.5,14){\includegraphics{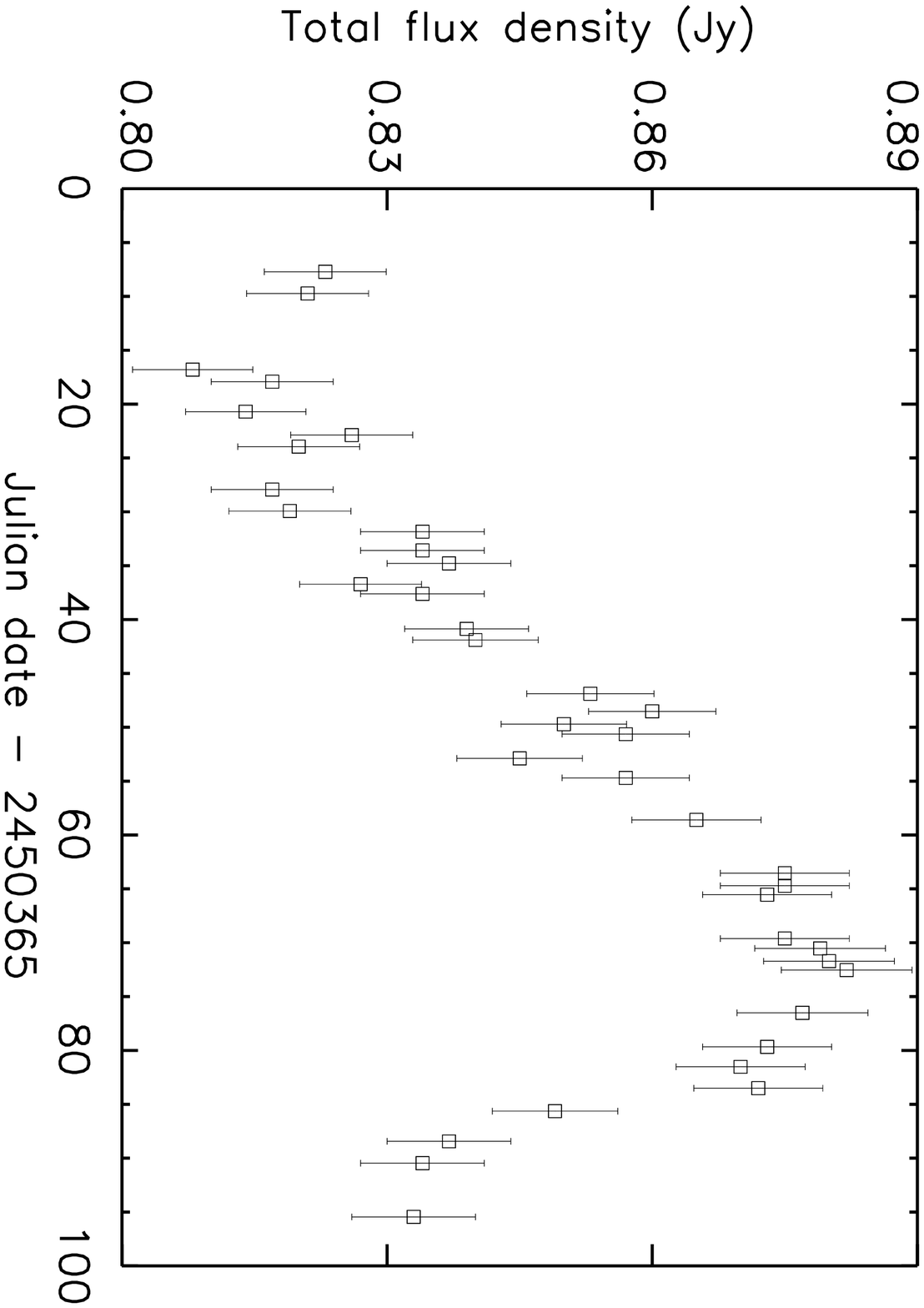}}
\put(10,21) {b)}
\put(18.5,14){\includegraphics{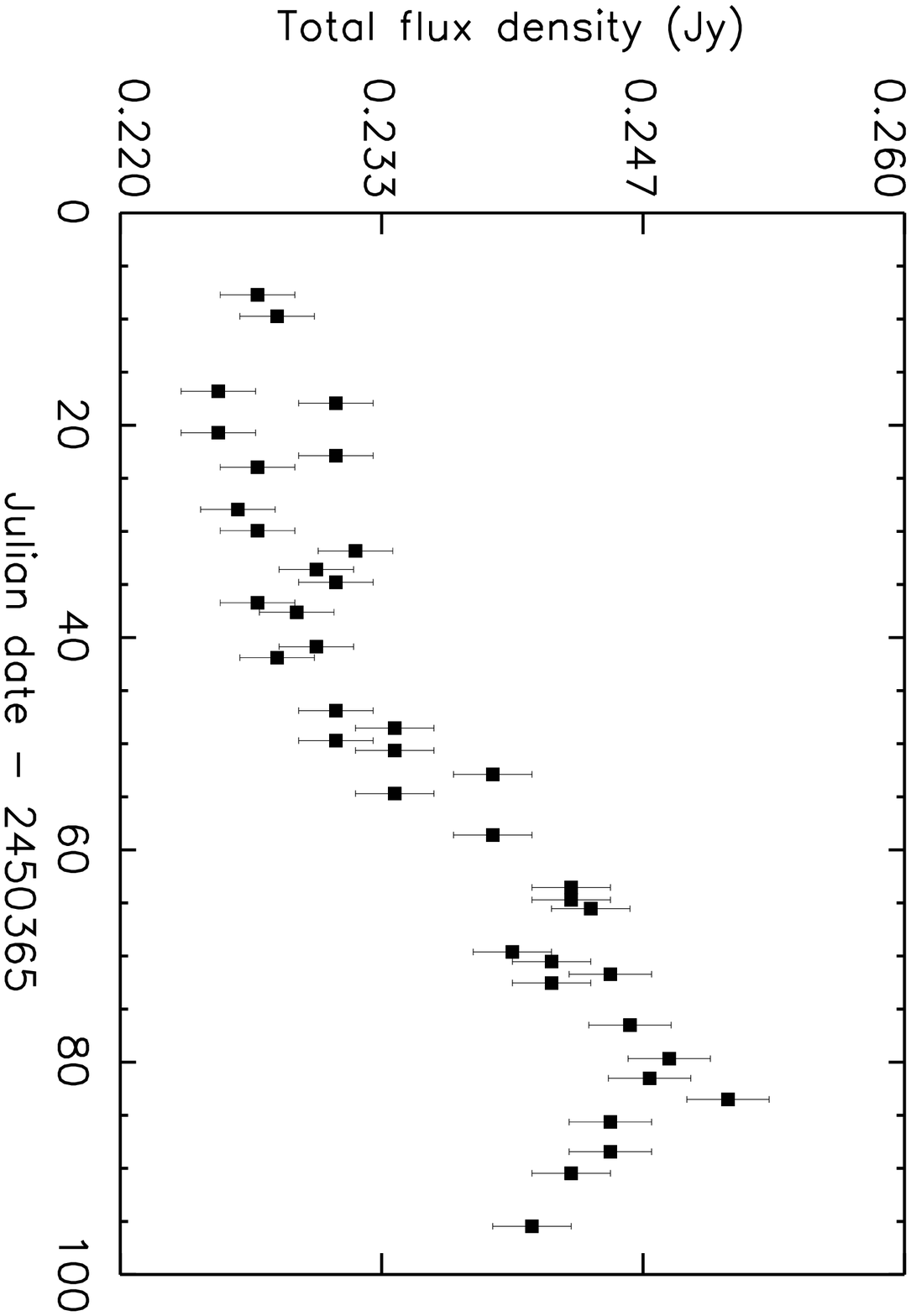}}
\put(1,14) {c)}
\put(9.5,7){\includegraphics{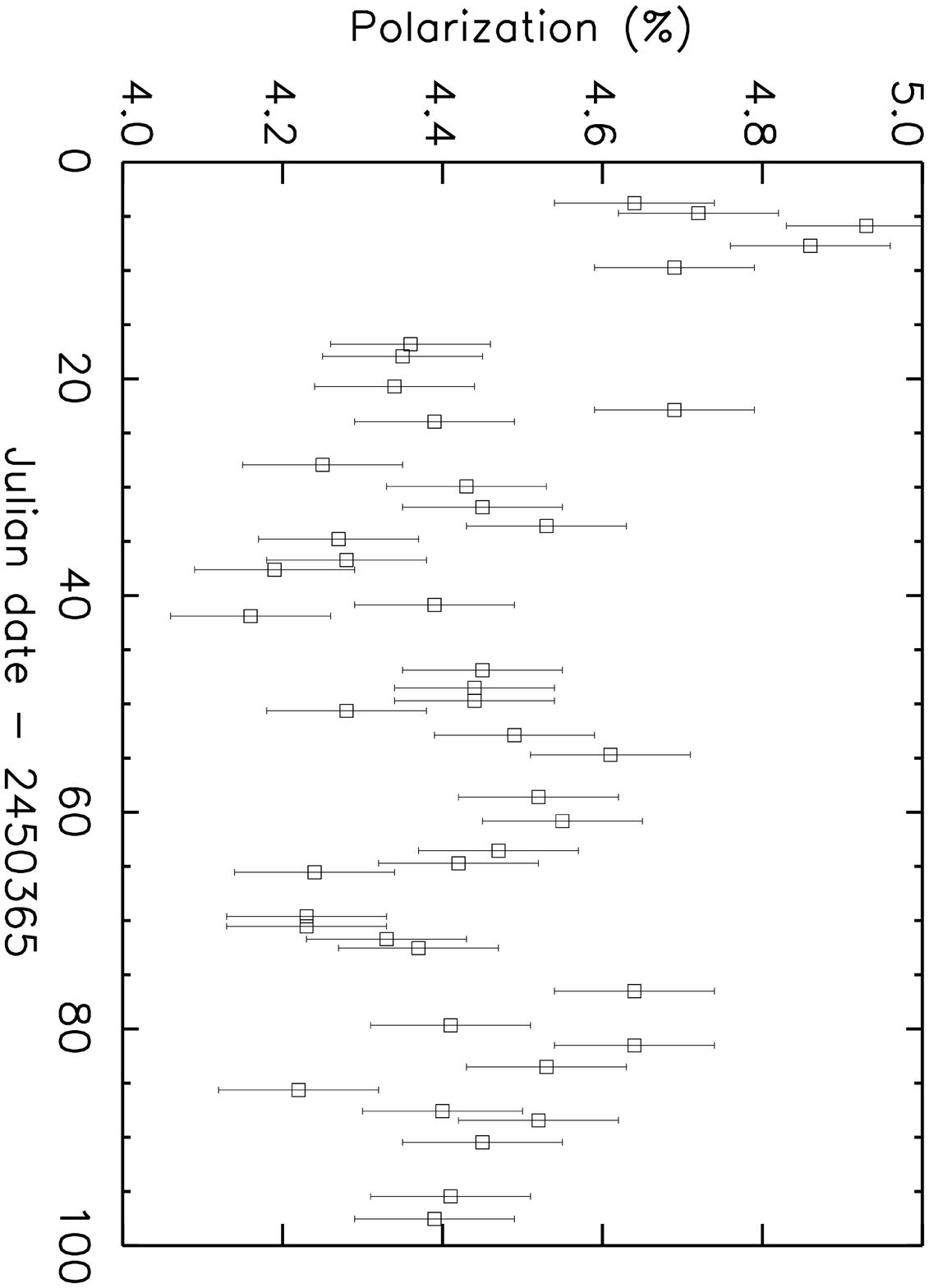}}
\put(10,14) {d)}
\put(18.5,7){\includegraphics{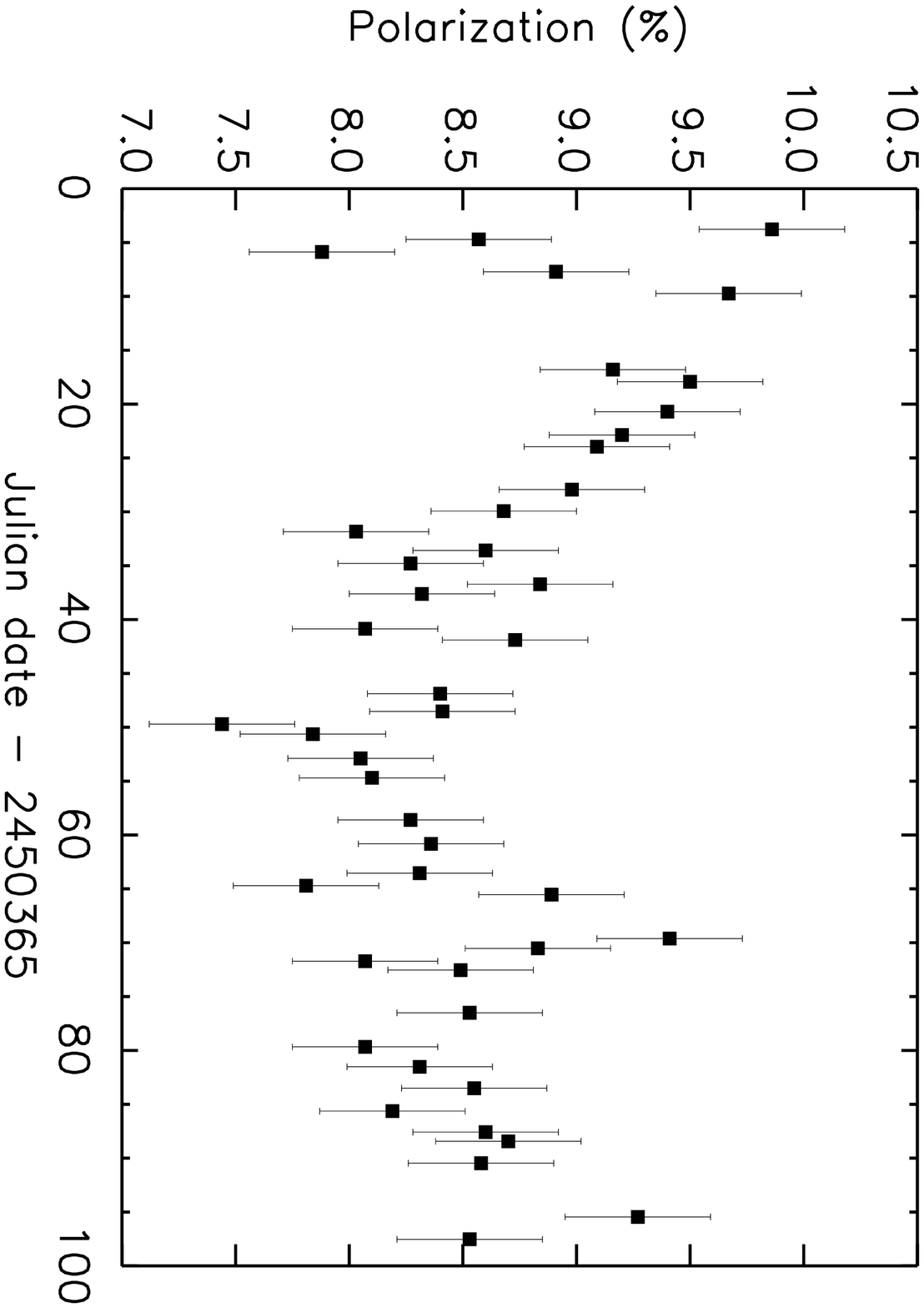}}
\put(1,7) {e)}
\put(9.5,0){\includegraphics{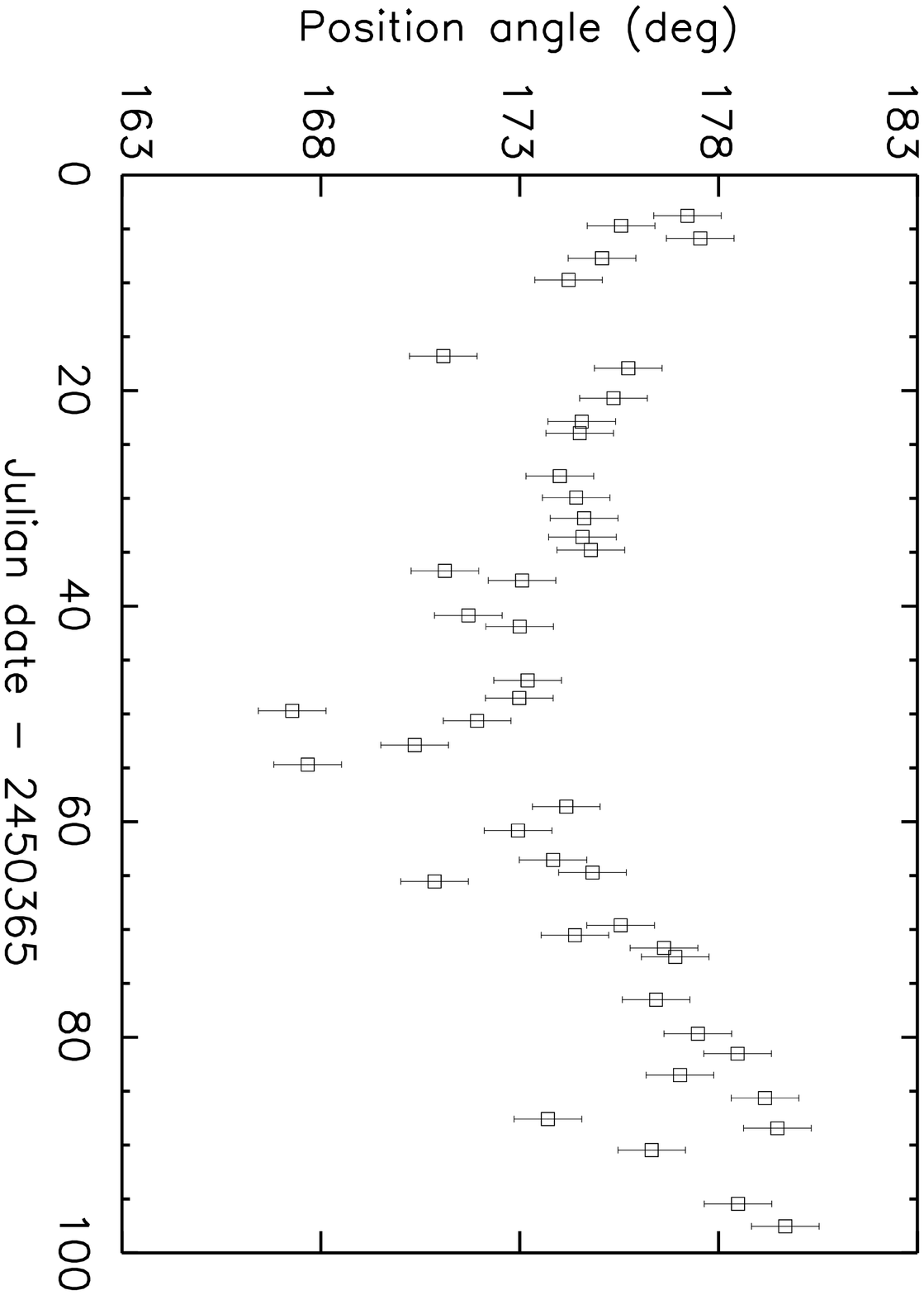}}
\put(10,7) {f)}
\put(18.5,0){\includegraphics{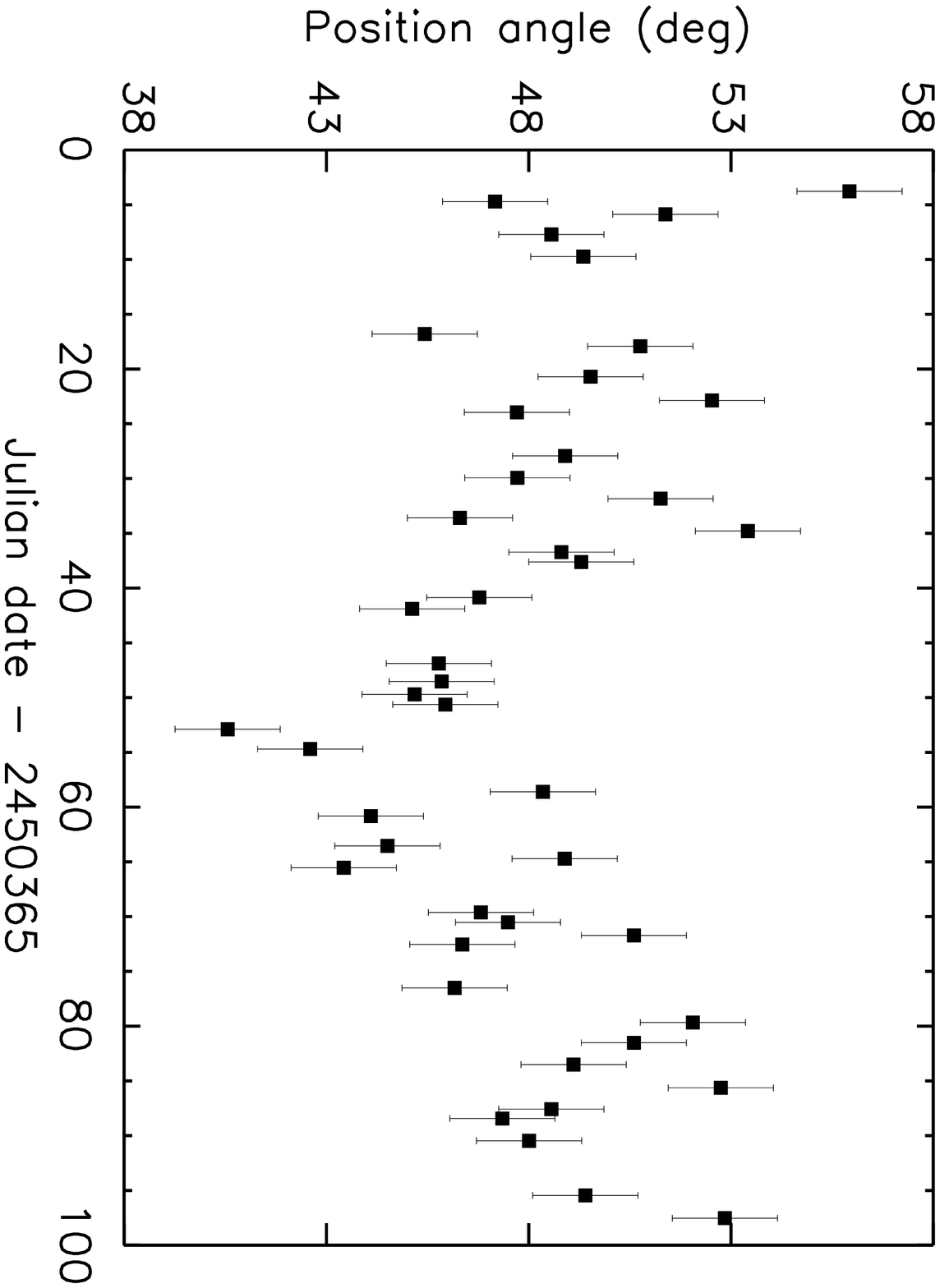}}
\end{picture}
\caption{8.4~GHz light curves. Top row: Total flux density of B0218+357, a) 
component A, b) component B. Middle row: Percentage polarization, c) 
component A, d) component B. Bottom row: Polarization position angle, e) 
component A, f) component B.}
\label{Xgraphs}
\end{center}
\end{figure*}

\begin{figure*}
\begin{center}
\setlength{\unitlength}{1cm}
\begin{picture}(20,15)(0,0)
\put(1,14) {a)}
\put(9.5,7){\includegraphics{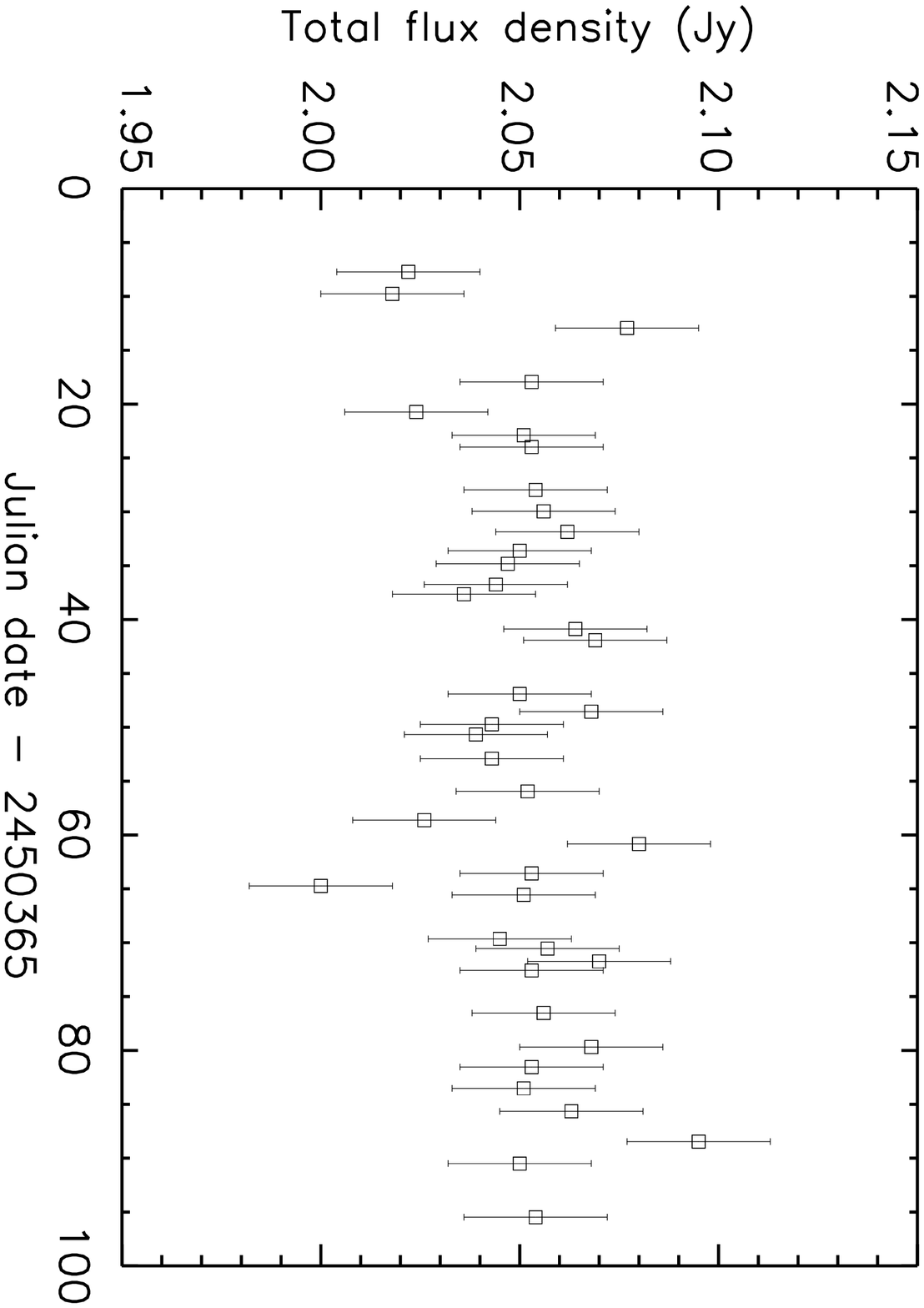}}
\put(10,14) {b)}
\put(18.5,7){\includegraphics{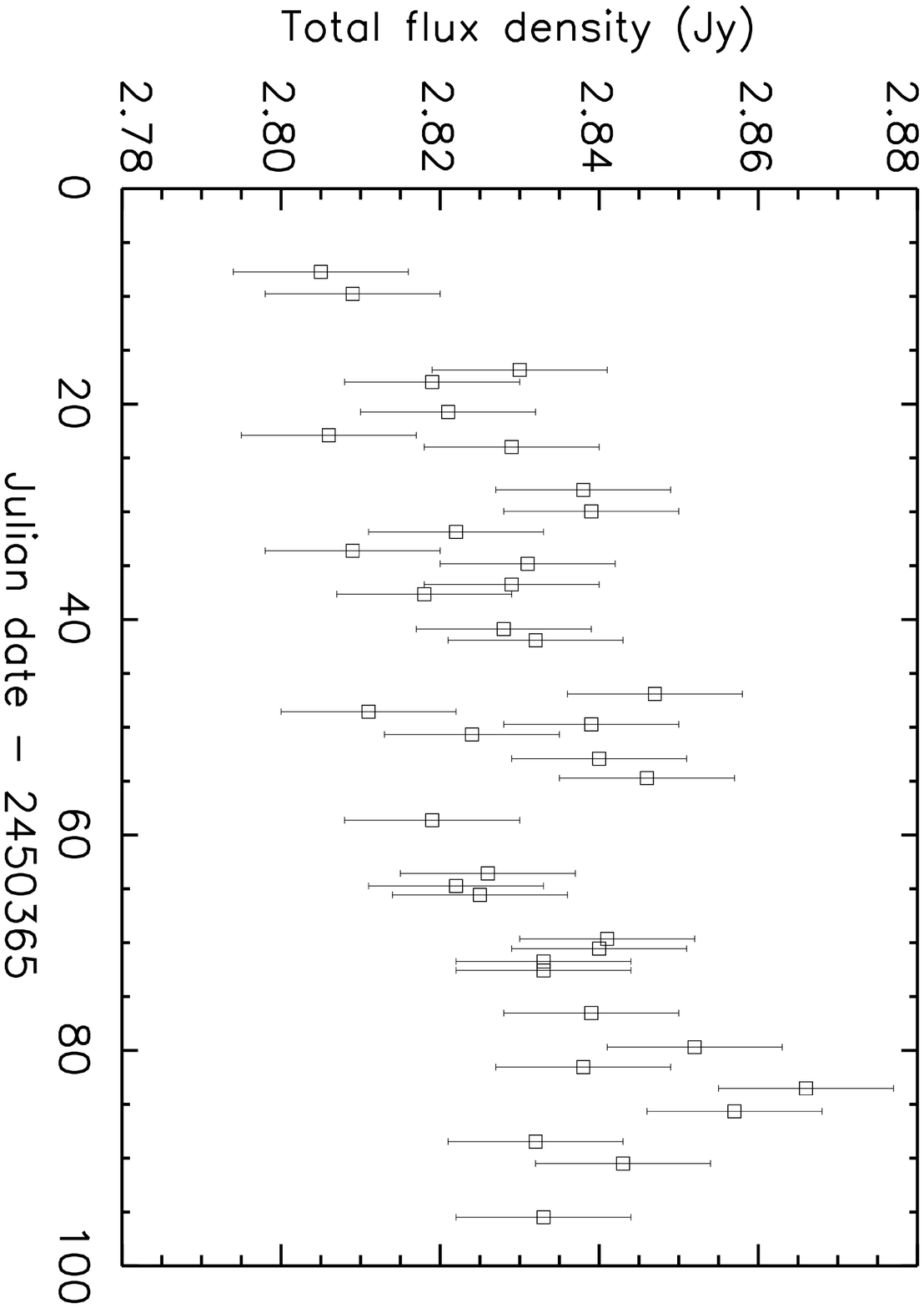}}
\put(1,7) {c)}
\put(9.5,0){\includegraphics{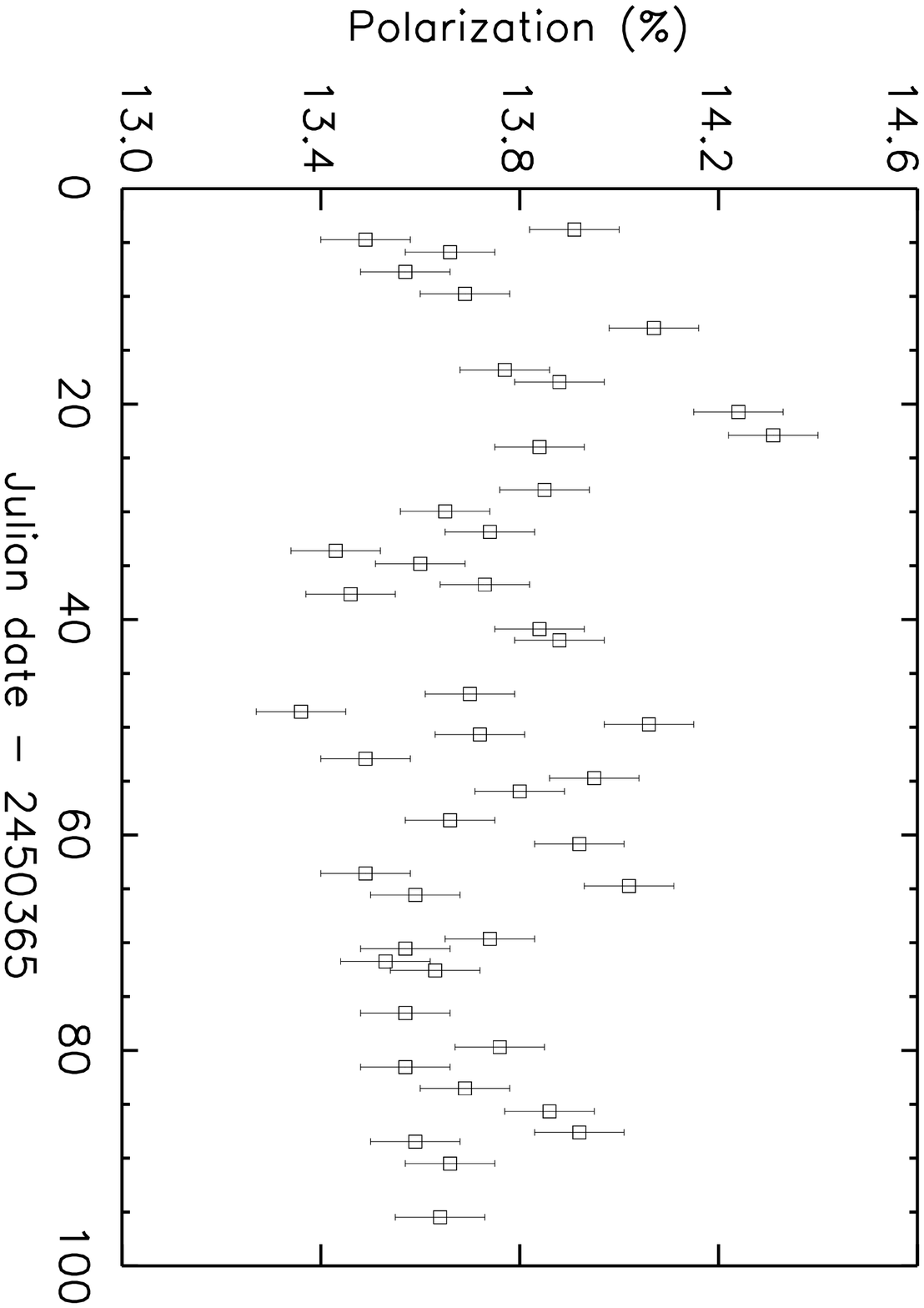}}
\put(10,7) {d)}
\put(18.5,0){\includegraphics{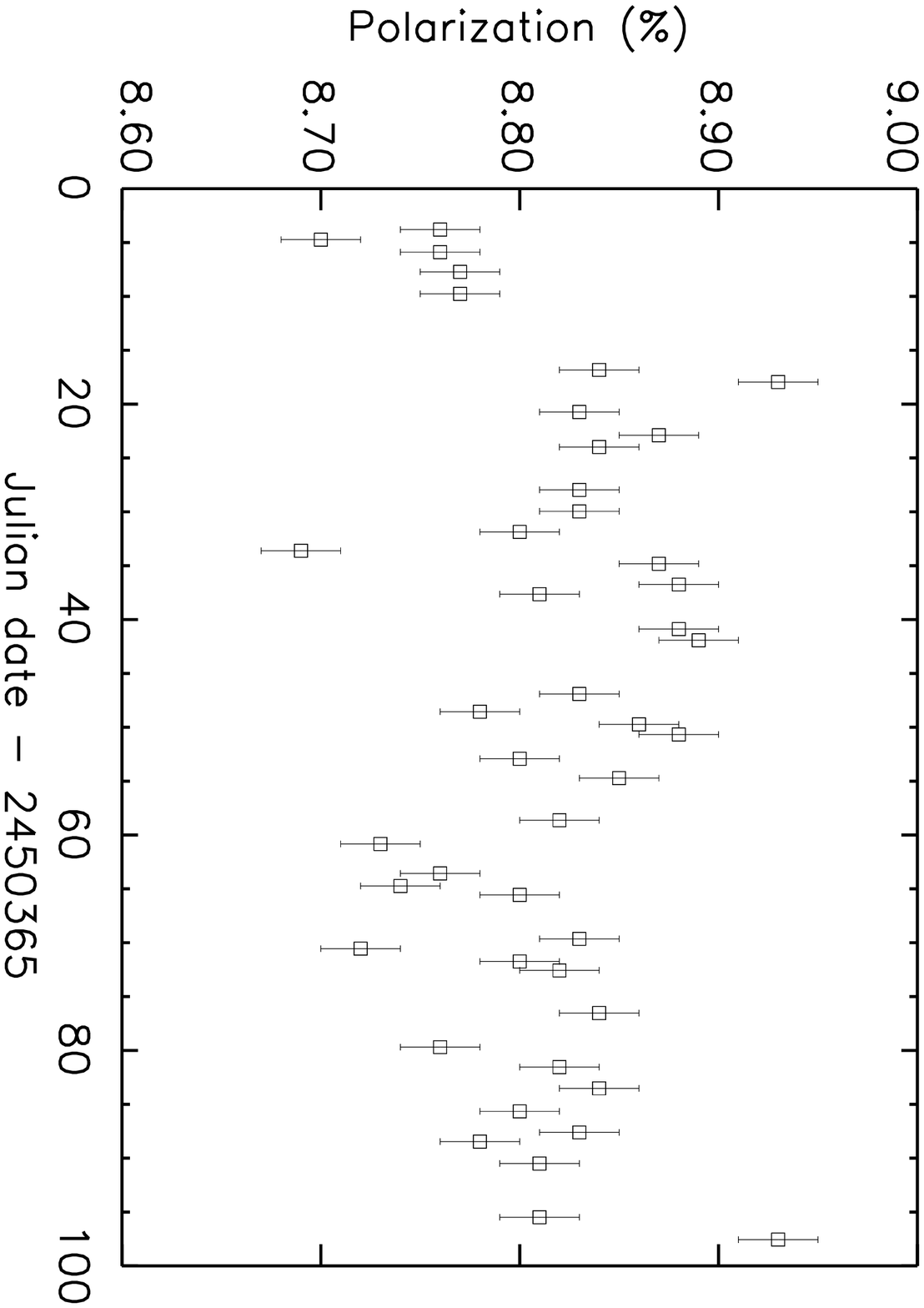}}
\end{picture}
\caption{Top row: Total flux density of the control source 3C119, a) 15~GHz, 
b) 8.4~GHz. Bottom row: Percentage polarization of 3C119, c) 15~GHz, 
d) 8.4~GHz. Note that we do not show the polarization position angle 
curves since 3C119 itself was used as the calibrator for the position 
angle of B0218+357.}
\label{3C119graphs}
\end{center}
\end{figure*}

The total flux density, percentage polarization and polarization position 
angle light curves for components A and B are shown in Fig.~\ref{Ugraphs} 
(15~GHz) and Fig.~\ref{Xgraphs} (8.4~GHz). As well as the light curves 
themselves we discuss in some detail the way in which the error bars on 
the points in each light curve are derived since these in the end determine 
the confidence we will have in the derived time delay. 

\subsection{Total flux density}

The variability and time delay signature are well illustrated by the 
15~GHz total flux density light curves, Fig.~\ref{Ugraphs} a) and b). 
The main trend in total flux density (seen at both frequencies) is a 
rise of about 10 per cent followed by a short plateau stage, and a 
sharp decline which is only just caught in the B component data. Also, 
prominent in the 15~GHz A component light curve are two dips in the 
flux density around days 20 and 30, both of which show up clearly in 
the B component data, but delayed by about 11 days relative to A. 
These features can be seen to a lesser extent in the 8.4~GHz light 
curves, Fig.~\ref{Xgraphs} a) and b). 

The variations in total flux density seen in the light curves for 
B0218+357 should be compared with those for the control source 3C119 
shown in Fig.~\ref{3C119graphs} a) and b). For both frequencies the 
rms scatter is $\la$1 per cent compared to the $\ga$10 per cent 
changes seen in B0218+357. Fig.~\ref{3C119graphs} appears to show 
that the total flux density of 3C119 increases by $\sim$1 per cent 
over the period of the observations. We do not believe this is a real 
change in 3C119, but an artifact arising from a gradual decrease in 
the flux density of 3C84 which was used as the flux density calibration 
source.  The total flux density of 3C84 has in fact been falling almost 
continuously since about 1983 (University of Michigan Interactive Radio 
Observatory Database, http://www.astro.lsa.umich.edu/obs/radiotel/umrao.html) 
and this manifests itself as the apparent small rise in the flux density of 
3C119 over the period of the monitoring observations. We tried normalising 
the data by dividing the B0218+357 total flux densities at each epoch by 
the corresponding 3C119 total flux density, but this had the effect of 
slightly increasing the scatter in the B0218+357 total flux densities and 
so has not been applied. 

The fact that normalizing the B0218+357 data by the 3C119 light curve does
not produce an improvement suggests that the residual scatter in the
3C119 data is due to observational, calibration and model fitting
uncertainties and is therefore a good measure of the flux density
measurement errors, both for 3C119 and B0218+357. We therefore use the
scatter on the 3C119 light curve to calculate an error bar for the
total flux density measurements at 15~GHz. If we fit a straight line
to the 3C119 data, the scatter around this is 18 mJy or 0.9 per
cent. As calibration errors are multiplicative, the error bar for each
epoch in Fig.~\ref{Ugraphs} a) and b) has been set
 equal to 0.9 per cent of the flux at that epoch, typical values of which
are 8 mJy for component A and 2 mJy for component B.

Using the same method to obtain the error bars at 8.4~GHz does not
work well. This is because the model fit to the B0218+357 data is so
much poorer (as measured by $\chi^{2}$) than that to 3C119. This
indicates that model fitting errors dominate over calibration errors
and the scatter in the 3C119 data at this frequency (11 mJy or 0.4 per
cent) is far too small to be a good estimate of the errors on the
B0218+357 8.4~GHz flux density points. In the end, we believe the
chi-squared minimization technique itself that we use to calculate the
delay between components A and B (described in Section~4.1) gives the
best estimate of the errors. This is because we expect a $\overline{\chi}^2$ 
of approximately unity if the errors on the data points are correctly chosen.
For a `moderately' good fit, the ${\chi}^2$ statistic has a mean of $\nu$ 
and a standard deviation of $\sqrt{2\nu}$ where $\nu$ is the number of 
degrees of freedom \cite{press92}. Therefore to be conservative we choose 
our errors such that ${\chi}^2$ at best fit is one standard deviation 
{\em below} $\nu$. For the B0218+357 total flux density data set which 
contains 38 epochs and is fitted with two parameters (delay and flux density 
ratio), this gives a required $\overline{\chi}^2$ equal to 0.76 which 
corresponds to errors of approximately 7 mJy for component A and 2 mJy 
for component B.

\subsection{Percentage polarization}

The 15~GHz polarization curves also contain variations over short time 
scales which provide an independent constraint on the time delay. 
The time series of percentage polarization at 15~GHz broadly consists of 
two features that stand out from the noise. Seen in component B, 
Fig.~\ref{Ugraphs} d), the first is a rise and fall of greater than 
1 per cent, taking place over a period of about 20 days, shortly 
after the beginning of the monitoring observations. The second is 
of similar magnitude and duration to the first and can be seen most 
clearly in Component A between days 70 and 90, Fig.~\ref{Ugraphs} c). 
This corresponds to the sharp drop in total flux density seen at both 
frequencies. The 8.4~GHz percentage polarization results, Fig.~\ref{Xgraphs} 
c) and d), are of a much poorer quality than those seen at 15~GHz, due to 
the combination of a poorer Gaussian model fit to the data and lower 
polarized flux density. Although there are some small variations that 
can be discerned, they are of such low significance compared to the 
other data sets that the 8.4~GHz polarization light curves have not been 
used in the main analysis below.

The percentage polarization light curves of 3C119 at 15~GHz and 8.4~GHz
are shown in  Fig.~\ref{3C119graphs} c) and d). Clearly the polarization 
calibration procedure works well though there are some residual systematic 
trends of $\sim$0.1 per cent. This suggests that perhaps the assumption 
that 3C84 has zero and non-variable polarization breaks down and that 
our derived polarization residuals are being contaminated by very small 
intra-day variations in 3C84. However, this will have a negligible effect 
on the B0218+357 polarization data as the variations in this source are 
much larger and occur at different times and on different time scales 
than those seen in the 3C119 light curves.

We use the rms difference between the measurements of percentage
polarizations obtained in the independent IFs in order to estimate the
error bars for each data point. These differences will arise from
thermal noise and model fitting errors and will, therefore, represent a
lower limit to the error bar for each component. We also correct each
IF difference for a constant offset that exists between the two IFs, the 
origin of which is unclear. At 15~GHz this offset is equal to 0.02 per cent 
for component A and 0.07 per cent for component B. At this same frequency the 
estimated errors are equal to 0.05 per cent for component A and 0.17 per 
cent for component B. 

\subsection{Polarization position angle}

By far the most prominent feature of the variation in polarization 
position angle at 15~GHz is the gradual rotation between days 35 and 
80 seen in both component A and B, Fig.~\ref{Ugraphs} e) and f), covering 
a range of about 10$^{\circ}$. As with percentage polarization at 8.4~GHz,
the polarization position angle variations at 8.4~GHz, Fig.~\ref{Xgraphs} 
e) and f), are poorly defined compared to those at 15~GHz and are therefore 
also excluded from the time delay analysis. As the polarization position 
angle of B0218+357 is calibrated by that of 3C119 we do not show plots of 
the 3C119 polarization position angle.

The error bars on the 15~GHz and 8.4~GHz polarization position angle data 
are calculated in the same manner as that described in Section~3.2 for the 
percentage polarization data. Error bars on the 15~GHz data are 0.58$^{\circ}$ 
and 0.80$^{\circ}$ for components A and B respectively.

\section{Time delay analysis}

\begin{figure}
\begin{center}
\setlength{\unitlength}{1cm}
\begin{picture}(5,6)(0,0)
\put(6.75,-0.5){\includegraphics{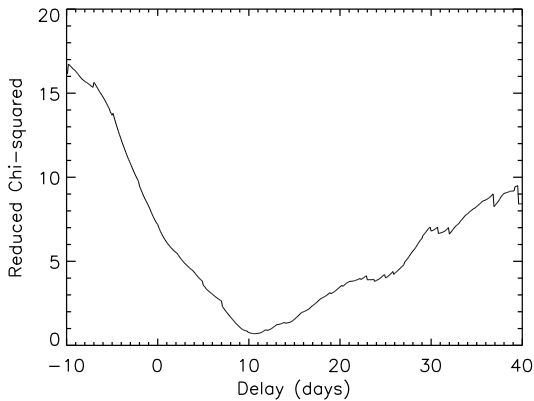}}
\end{picture}
\caption{$\overline{\chi}^2$ against delay for total flux density, 15~GHz.}
\label{examplechi}
\end{center}
\end{figure}

\subsection{The chi-squared method}

In order to quantify the time delay between components A and B, a 
chi-squared analysis has been performed on each data set. Each point 
in one light curve, $A(t_i)$, is shifted by some delay $\tau$ (between 
$-10$ and 40 days in steps of 0.1 of a day) and paired with a linearly 
interpolated value from the other, $B(t_i)$, interpolation being required 
due to the uneven and sparse sampling of data. All points for which 
$t_i + \tau > t_N$ and $t_i - \tau < t_1$, where $N$ is the total number 
of epochs, are excluded. As there is no reason to prefer interpolation 
in one time series over the other, the analysis is then repeated, but 
by shifting $B(t_i)$ by $-\tau$, interpolating in $A(t_i)$ and then 
averaging the results of the two passes. Since the measured chi-squared 
is inversely proportional to the number of overlapping data points, delays 
for which the overlap is small are weighted down relative to those for which 
the overlap is large. Furthermore, as well as shifting the data temporally,
the data are also scaled along the y-axis in order to produce the best
alignment of the A and B light curves. Scalings are determined by first
calculating the ratio (total flux density and percentage polarization) 
or difference (polarization position angle) between each A and B pair
in the unshifted data set and averaging over the entire light curve. A 
range of scalings that bracket this figure are then determined and sampled
in steps of 0.01 for total flux density and percentage polarization and 
steps of 0.1 for polarization position angle. For each trial value of the
delay, the value of this scaling is found which minimizes chi-squared, so
calculating the flux density ratio/de-polarization/differential 
rotation between A and B. An example plot of $\overline{\chi}^2$ against 
delay is shown in Fig.~\ref{examplechi} for total flux density, 15~GHz. 

\begin{figure*}
\begin{center}
\setlength{\unitlength}{1cm}
\begin{picture}(20,15)(0,0)
\put(1,13.5) {a)}
\put(-0.4,14.5){\includegraphics{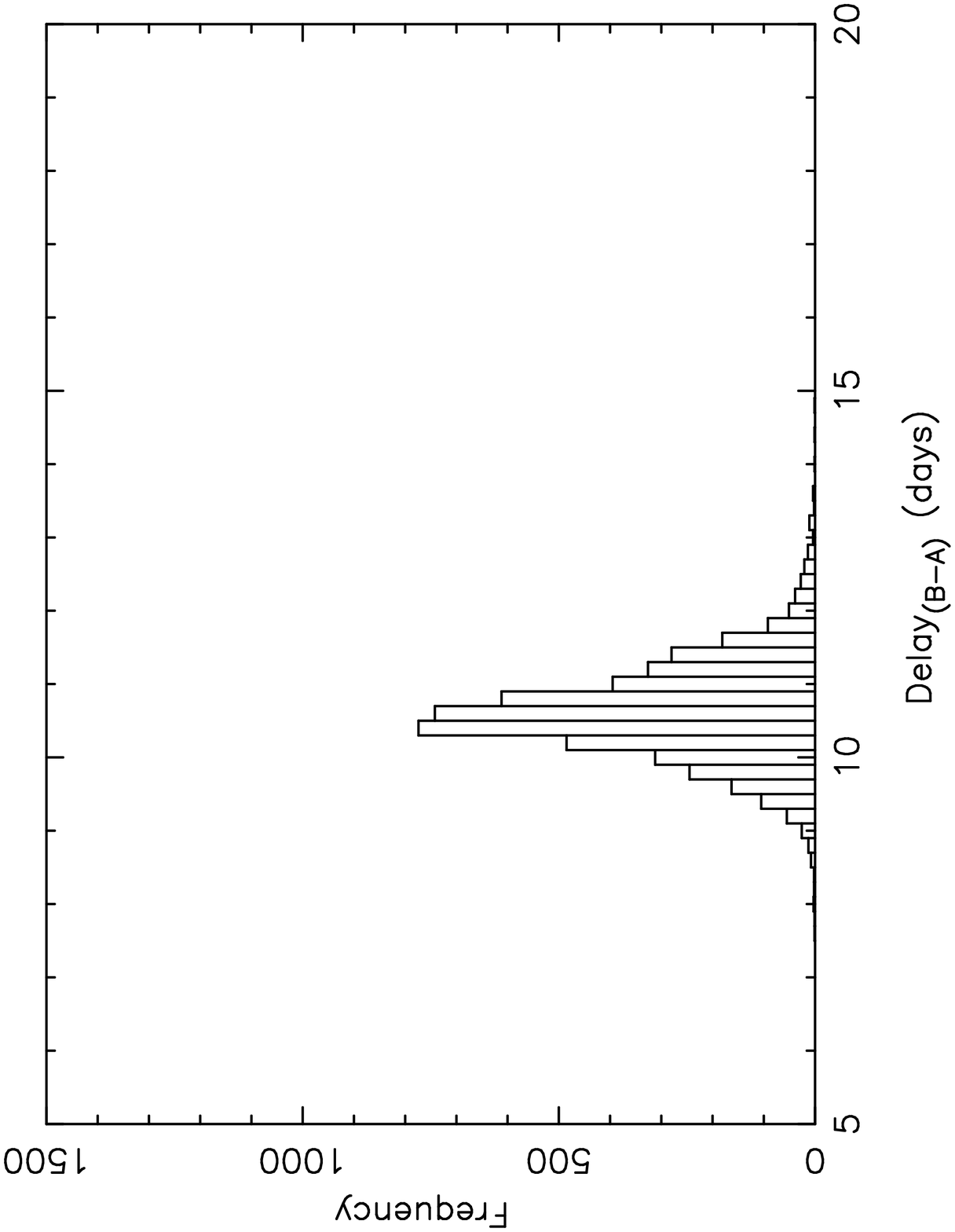}}
\put(10,13.5) {b)}
\put(8.6,14.5){\includegraphics{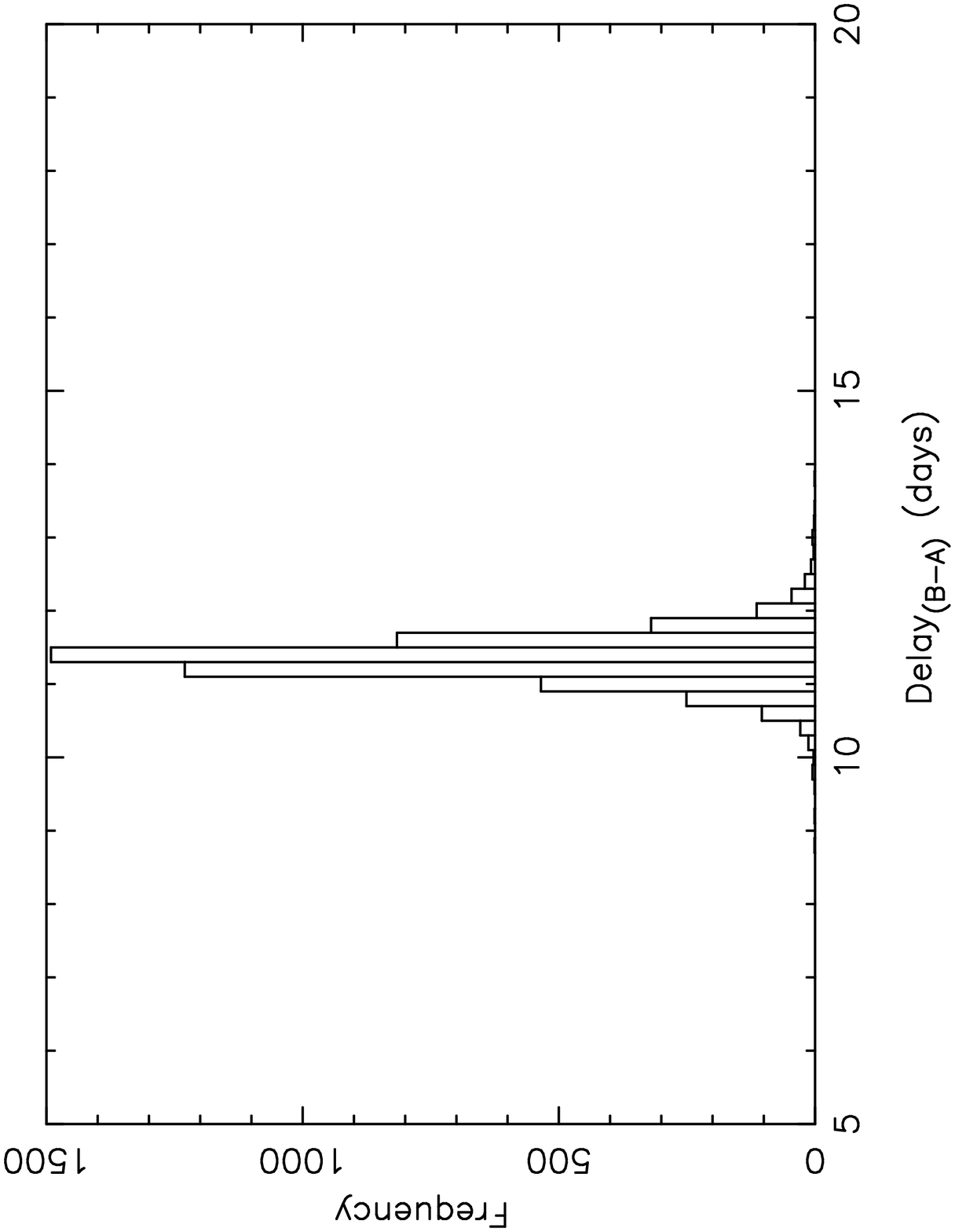}}
\put(1,6.5) {c)}
\put(-0.4,7.5){\includegraphics{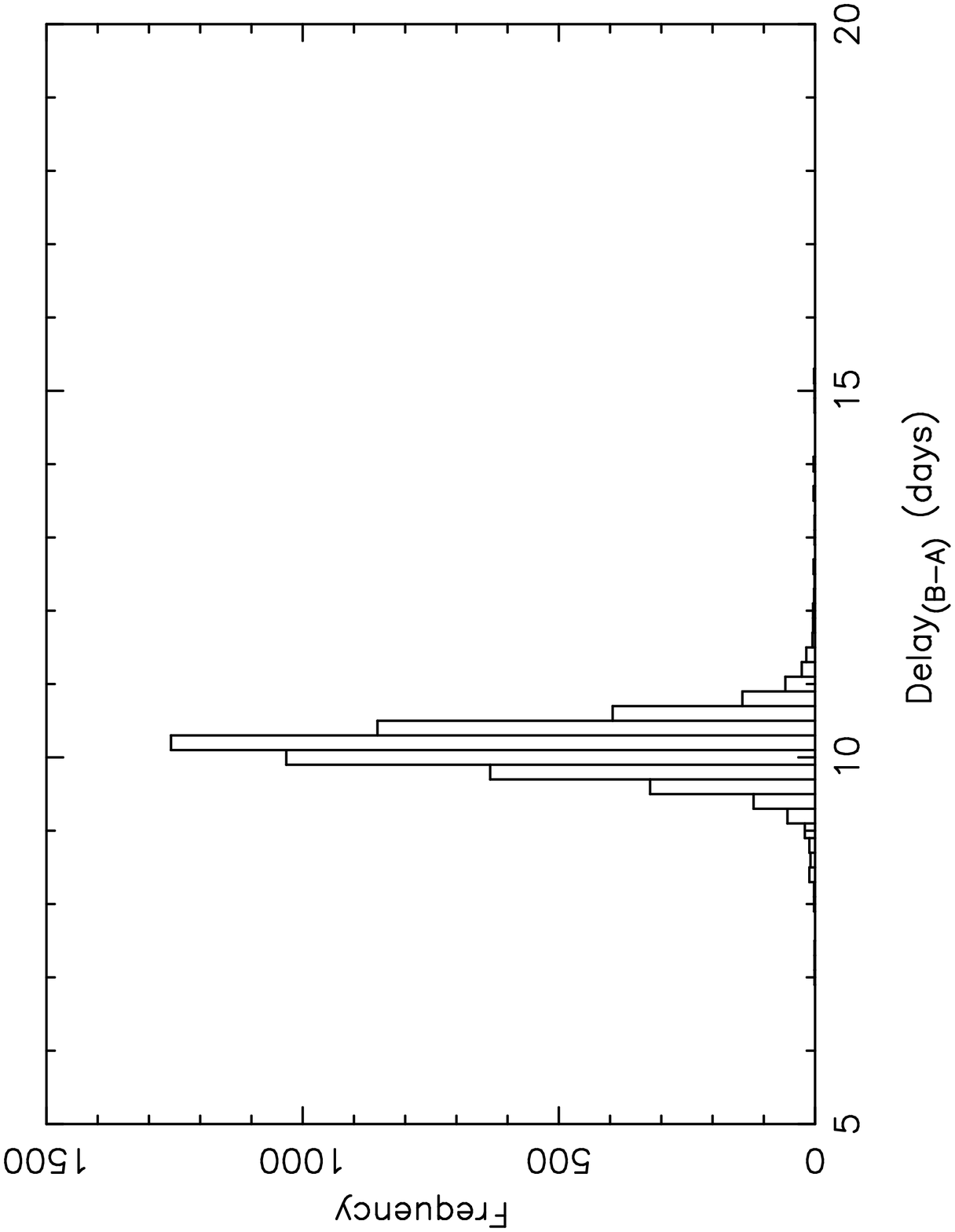}}
\put(10,6.5) {d)}
\put(8.6,7.5){\includegraphics{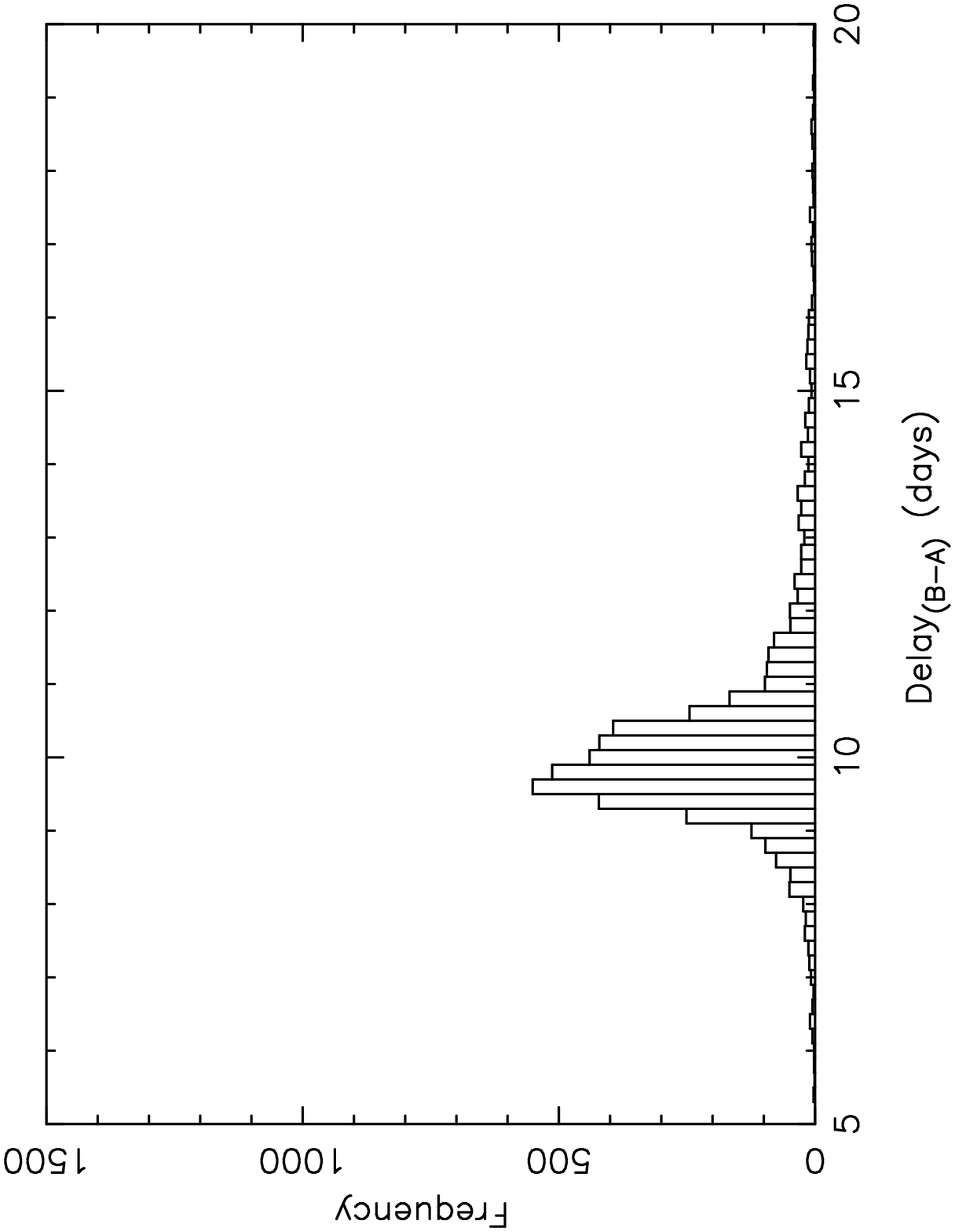}}
\end{picture}
\caption{Frequency histograms of delay (5,000 realizations) for a) Total 
flux density, 15~GHz, b) Percentage polarization, 15~GHz, c) Polarization 
position angle, 15~GHz and d) Total flux density, 8.4~GHz. Delays are 
binned in 0.2 day intervals.}
\label{histograms}
\end{center}
\end{figure*}

As well as the chi-squared analysis described above we have also used 
a variety of other methods to determine the time delay, all of which 
have been used in the past for the analysis of B0957+561 monitoring 
data. We have calculated the cross-correlation function of the A and B 
light curve data, the discrete correlation function \cite{edelson88}
and the dispersion measure $D^2_2$ \cite{pelt96}, the last two specifically
taking into account the fact that the A and B time series consist of 
unevenly sampled data and hence do not require interpolation. These 
techniques all give results consistent with those found using a chi-squared 
minimization and so in this paper we choose to concentrate on the results 
found from the chi-squared analysis.

\subsection{Error on the delay - Monte-Carlo simulations}

\begin{table*}
\begin{center}
\caption{Best-fit delays and magnitude scalings with associated 68.3 per 
cent confidence limits from the chi-squared analysis described in Section~4.1 
and the Monte-Carlo simulations described in Section~4.2, for all data sets. 
$^{\dagger}$A/B, $^{\ddagger}$A$-$B.}
\begin{tabular}{cccc}
 & Delay (days) & Flux density ratio$^{\dagger}$ & $\overline{\chi}^2$ 
\vspace{0.2cm} \\

Total flux density (15~GHz) & $10.6^{+0.7}_{-0.5}$ 
& $3.73^{+0.01}_{-0.01}$ & 0.70 \vspace{0.2cm} \\

 & Delay (days) & De-polarization fraction$^{\dagger}$ \vspace{0.2cm} 
& $\overline{\chi}^2$  \\ 

Percentage polarization (15~GHz) & $11.4^{+0.3}_{-0.3}$ 
& $0.92^{+0.00}_{-0.01}$ & 2.26 \vspace{0.5cm} \\

 & Delay (days) & Differential rotation ( $^{\circ}$)$^{\ddagger}$ 
& $\overline{\chi}^2$ \vspace{0.2cm} \\

Polarization position angle (15~GHz) & $10.2^{+0.3}_{-0.4}$ 
& $-15.4^{+0.2}_{-0.2}$ & 1.14 \vspace{0.4cm} \\

 & Delay (days) & Flux density ratio$^{\dagger}$ & $\overline{\chi}^2$ 
\vspace{0.2cm} \\

Total flux density (8.4~GHz) & $10.1^{+1.4}_{-0.7}$ & $3.57^{+0.01}_{-0.01}$ 
& 0.76 \vspace{0.5cm} \\

\end{tabular}
\label{delaytable}
\end{center}
\end{table*}

It is important to have an objective assessment of the confidence
limits on the derived time delays. To do this we have adopted a 
Monte-Carlo approach taking into account both the uncertainty introduced by
the measurement errors and those introduced by the gaps in the
sampling of the light curves.  We do this independently for each
data set since the errors are different in each case.  

The first step is to try to create the best estimate of the true light 
curve of the object by combining the A and B light curves, making use of 
our estimate of the time delay and by scaling with our estimate of the
image flux density ratio, de-polarization fraction or position angle
difference. Examples of combined light curves are shown in 
Fig.~\ref{composites}. The second stage is to re-sample with 
replacement\footnote{A sampling interval once chosen is returned to the 
distribution.} the light curves at intervals chosen randomly from the 
distribution of real sampling intervals used for the original VLA 
observations. We choose the new sampling intervals so that the total 
length of the new light curve does not exceed that of the original. We 
also constrain its total length to be within 2.5 days of the original. 
Thus we ensure that all simulated light curves are of comparable length. 
Using these sampling intervals, the combined light curve is re-sampled 
twice, once with no delay and once using the delay estimated from the 
real data. Note that, by re-sampling the combined light curve without 
any smoothing, i.e. using linear interpolation, we take into account 
automatically much of the spread in the real data introduced by the 
effects of calibration errors etc. The resulting light curve is then 
{\em de}-magnified, {\em re}-polarized or {\em re}-rotated by the 
appropriate amount used originally to create the combined light curve. 
New realizations of the A and B light curves are thus created. The samples 
forming the light curves for each epoch are then perturbed by adding random 
noise drawn from a normal distribution with a certain standard deviation. 
This standard deviation is equal to the error bars on the points, the 
derivations of which are described in Section~3. From this perturbed set of 
light curves the best-fit delay and scaling is then calculated in exactly 
the same way as for the real data. This process is repeated many times, 
every time re-sampling the simulated data set. In this way frequency 
histograms are produced which should correspond to the error distribution 
of the measured delay. If anything this approach {\em over-estimates} the 
uncertainty in the time delay since there is some `double counting' of the 
errors; they contribute to the spread in the combined light curve which is 
then resampled, and again when the resampled light curves are perturbed 
according to the errors on the points.

Frequency histograms for the delay data are shown in Fig.~\ref{histograms}. 
Each is the result of 5,000 realizations and is plotted with a bin width of 
0.2 day. The distributions for both delay and magnitude scaling are highly 
non-Gaussian and so cannot be described completely by their full width at 
half maximum, which would correspond to a 1$\sigma$ confidence interval.
Note that each histogram has been plotted on the same axes thus clearly
illustrating the relative merits of each data set.

The time delays, magnitude scalings, associated 68.3 per cent confidence 
limits and $\overline{\chi}^2$ for each data set are shown in 
Table~\ref{delaytable} where the delay or magnitude scaling is equal to 
the median of the Monte-Carlo distribution and the confidence limits are 
calculated by counting inwards from the tails of the distributions until 
each tail contains 15.85 per cent of the total distribution.

These results show that in general our derived time delays are compatible 
at 1$\sigma$. Although the percentage polarization and position angle
data sets are not formally compatible at this confidence level, both 
are individually consistent with the time delays determined from the other 
data sets. 

\subsection{The final delay estimate}

\begin{figure}
\begin{center}
\setlength{\unitlength}{1cm}
\begin{picture}(5,6)(0,0)
\put(6.75,-0.5){\includegraphics{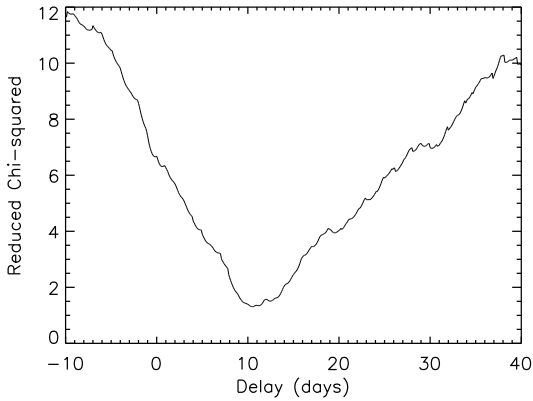}}
\end{picture}
\caption{$\overline{\chi}^2$ against delay for the simultaneous chi-squared
minimization.}
\label{simchi}
\end{center}
\end{figure}

In order to arrive at a final best estimate of the time delay, the 
results from each of the four light curves (total flux density, 15~GHz 
and 8.4~GHz, percentage polarization and position angle, 15~GHz) 
are combined. We utilize a `simultaneous chi-squared minimization' 
technique which makes use of all the information contained in the 
$\overline{\chi}^2$ against delay plots by summing the $\overline{\chi}^2$ 
found at each delay for each dataset. The minimum of the resulting 
distribution (see Fig.~\ref{simchi}) is that which is most consistent 
with all data sets and which represents our best-fit delay. This is equal 
to 10.5 days.

\begin{figure}
\begin{center}
\setlength{\unitlength}{1cm}
\begin{picture}(5,6)(0,0)
\put(-1.9,7){\includegraphics{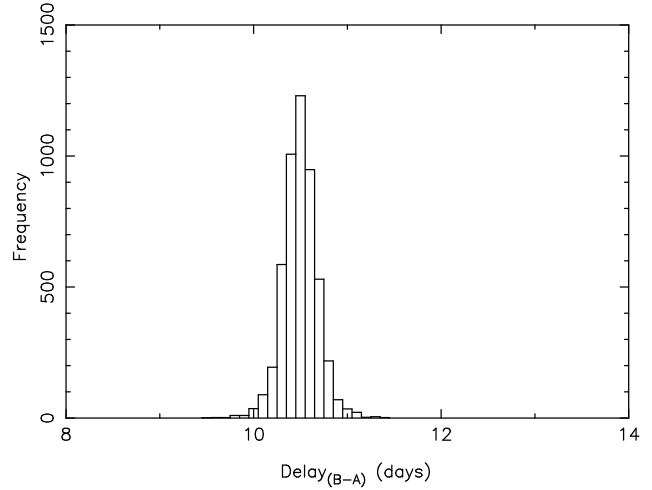}}
\end{picture}
\caption{Frequency histogram illustrating the results of the simultaneous
chi-squared minimization. Delays are binned in 0.1 day intervals.}
\label{finalhistogram}
\end{center}
\end{figure}

To establish the error on this delay, we perform a Monte-Carlo simulation 
similar to that described in Section~4.2, forming a combined light curve 
from each of the four light curves, but using the best-fit delay found in 
the simultaneous minimization for each. These are then re-sampled as before 
to produce realizations of the original light curves and another delay found 
by simultaneously minimizing chi-squared. This is repeated for a total of 
5,000 realizations. The result from this simulation is plotted as a frequency 
histogram in Fig.~\ref{finalhistogram} and the confidence limits given in 
Table~\ref{finaldelay}. At 95 per cent confidence the error on the time 
delay between images A and B in B0218+357 is $\pm0.4$ days.

\begin{table}
\begin{center}
\caption{Allowed ranges in the combined delay for different confidence levels
- simultaneous method.}
\begin{tabular}{cc}
Confidence (per cent) & Delay (days) \\ \hline
68.3 & 10.3 -- 10.7 \\
90   & 10.2 -- 10.8 \\
95   & 10.1 -- 10.9 \\
99   & 10.0  -- 11.1 \\
\end{tabular}
\label{finaldelay}
\end{center}
\end{table}

As a cross-check on the simultaneous chi-squared method we take a weighted 
average of the 5,000 delays which comprise the results of the individual 
Monte-Carlo simulations (plotted in Fig.~\ref{histograms}). The delays are
averaged realization by realization, weighting each by the inverse square 
of the full width of the 1$\sigma$ error bar (see Table~\ref{delaytable}). 
Though the delay distributions obtained using this method are skewed to higher 
delays, the medians of the distributions using both methods are compatible 
at 1$\sigma$. Our conclusion is therefore that at 95 per cent confidence
the best estimate of the delay is \boldmath${10.5\pm0.4}$\unboldmath
\,{\bf days}.

\section{Discussion and Conclusions}

\begin{figure*}
\begin{center}
\setlength{\unitlength}{1cm}
\begin{picture}(20,15)(0,0)
\put(1,14) {a)}
\put(9.5,7){\includegraphics{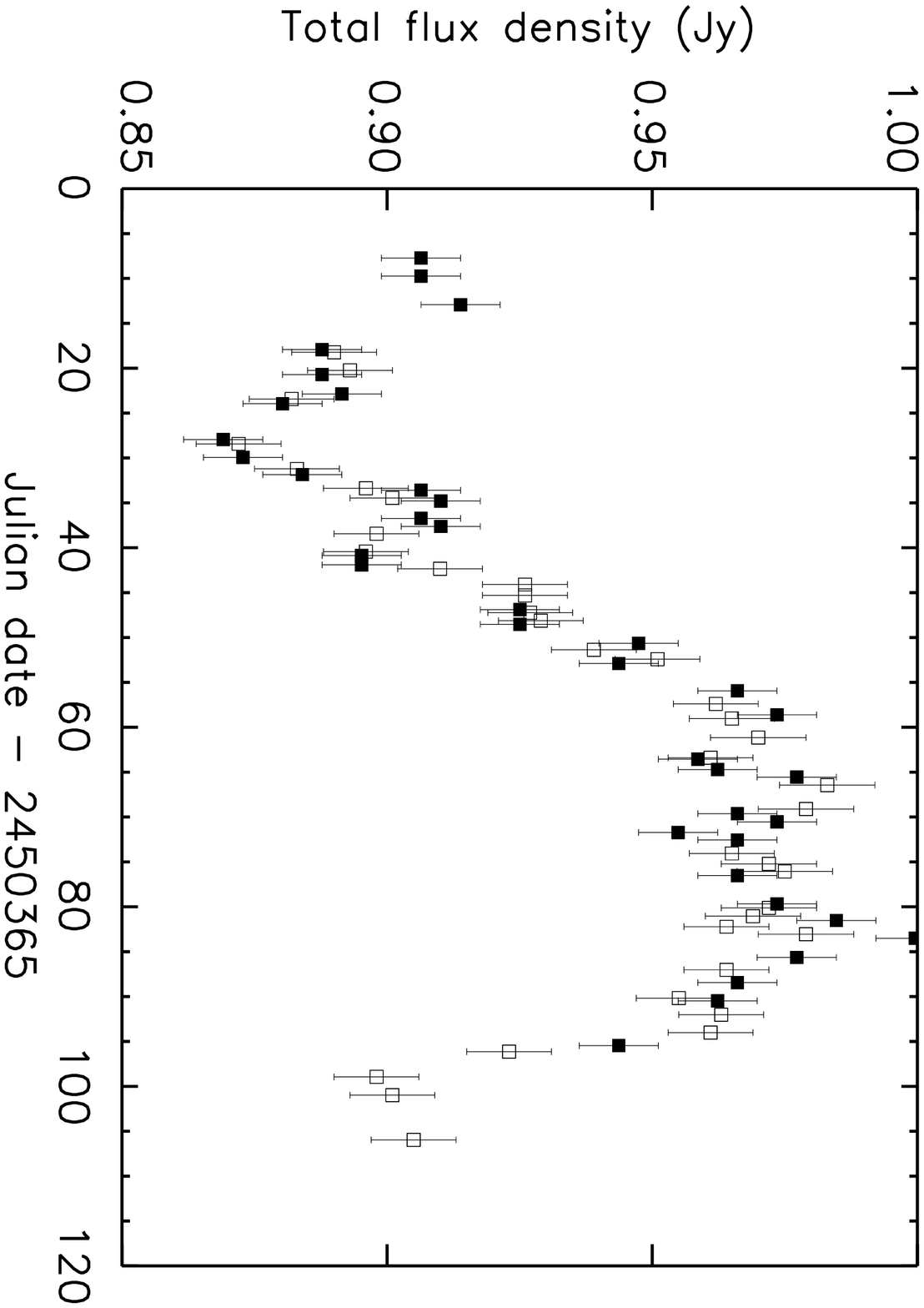}}
\put(10,14) {b)}
\put(18.5,7){\includegraphics{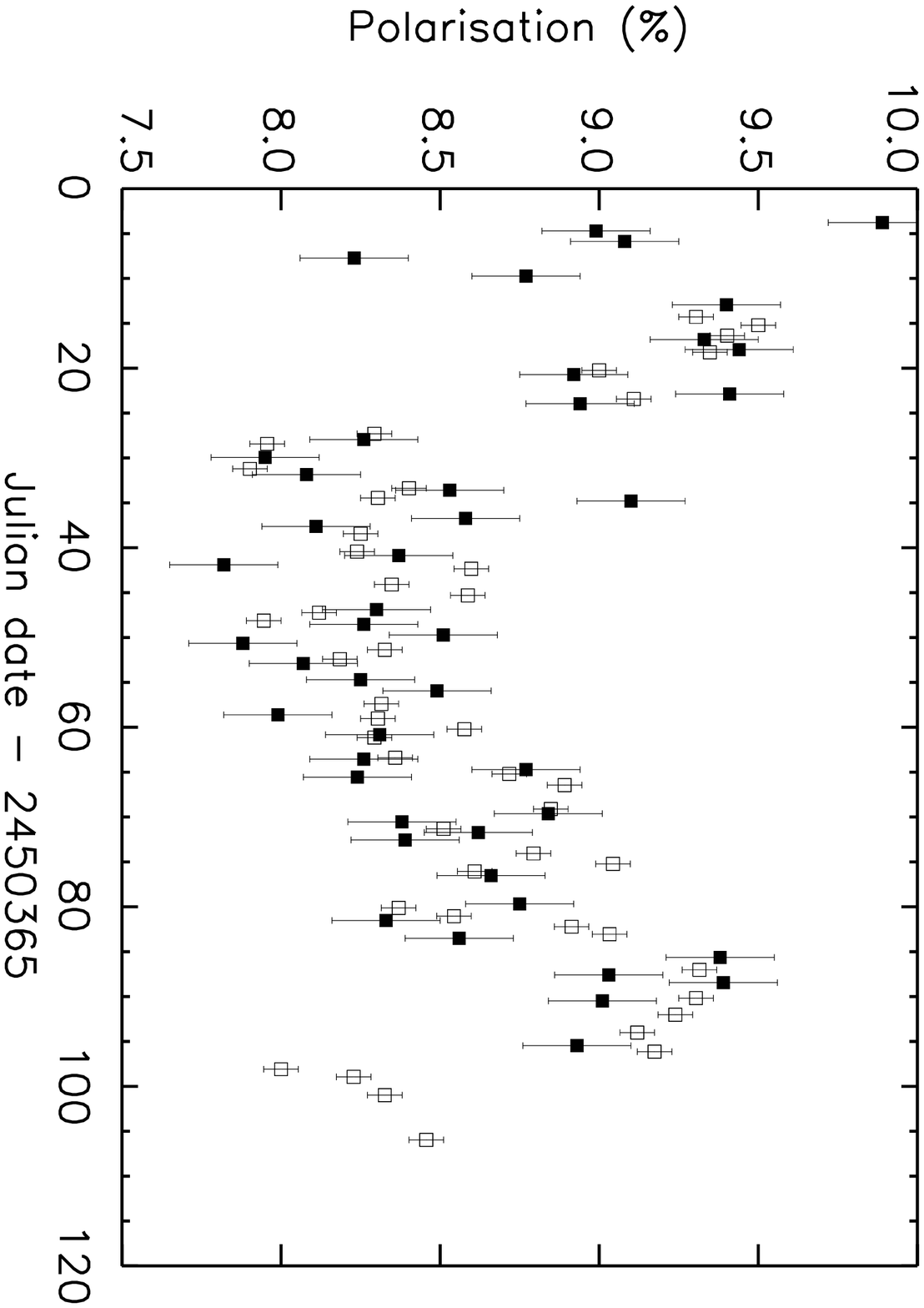}}
\put(1,7) {c)}
\put(9.5,0){\includegraphics{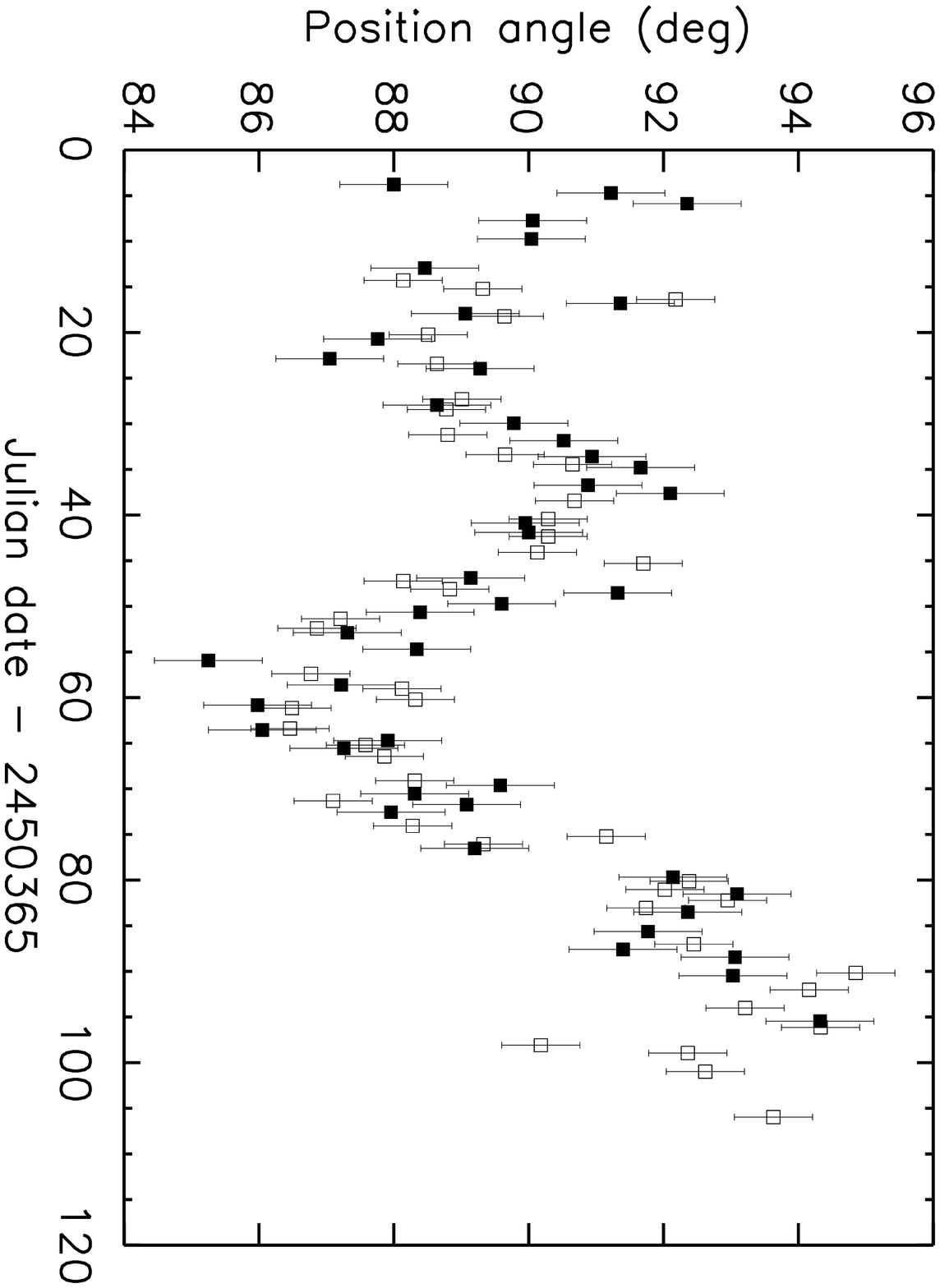}}
\put(10,7) {d)}
\put(18.5,0){\includegraphics{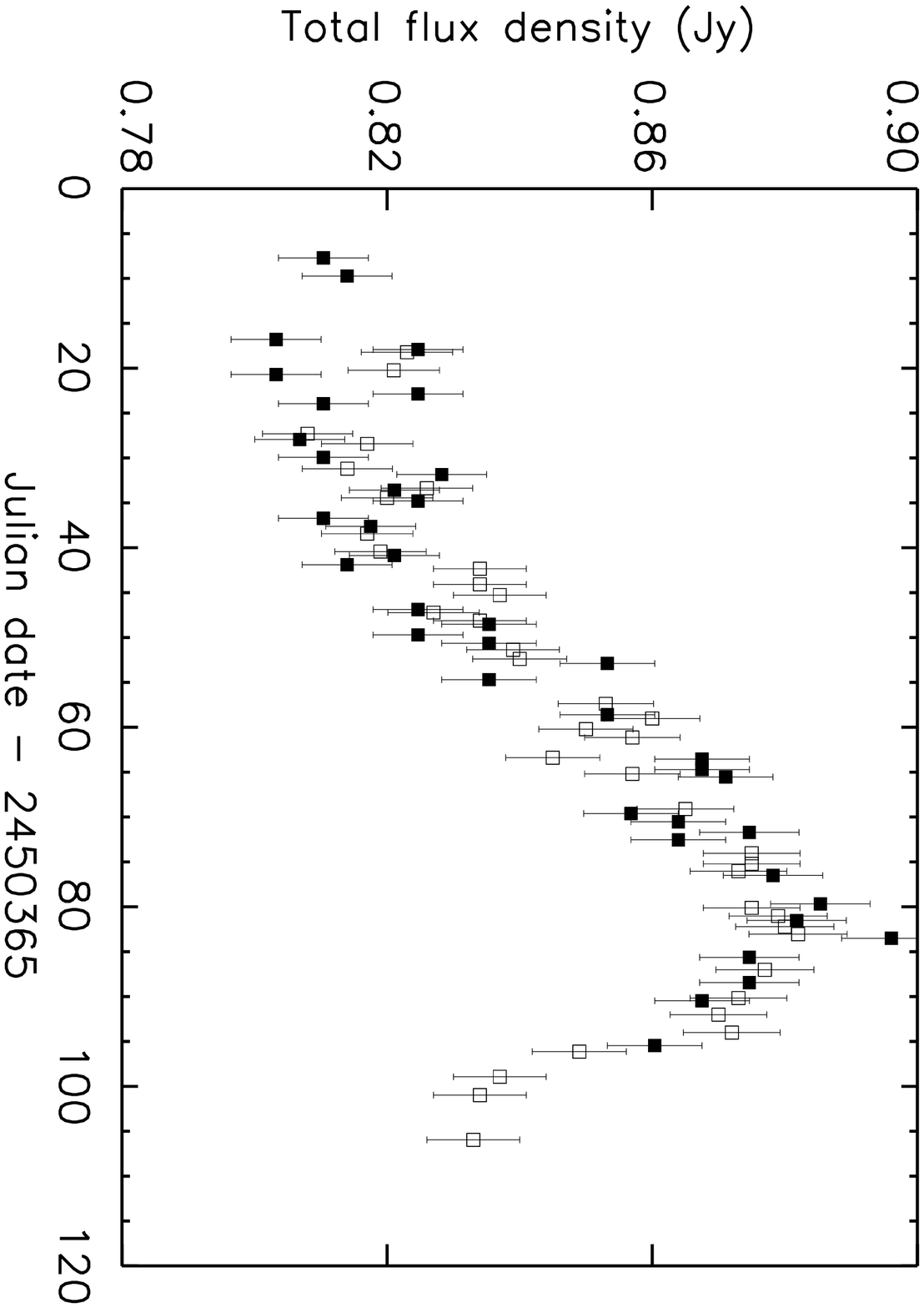}}
\end{picture}
\caption{Combined light curves. a) Total flux density, 15~GHz, b) 
Percentage polarization, 15~GHz, c) Polarization position angle, 15~GHz, 
d) Total flux density, 8.4~GHz. As in Figs~\ref{Ugraphs} and~\ref{Xgraphs}, 
component A and B measurements are represented by open and filled squares 
respectively.}
\label{composites}
\end{center}
\end{figure*}

Using new VLA observations of B0218+357, a value for the time delay 
between the variations of the two components in the gravitational lens 
system has been measured that is consistent with, and which represents an 
order of magnitude improvement on, the previous estimate of $12\pm3$ days 
\cite{liz96}. However, there is still room for improvement and 
observations covering a longer time period and with more frequent sampling 
than those presented in this work will allow the error to be further reduced.

Using the best estimate of the delay with the best-fit magnitude scalings, 
composite light curves are created by delaying each component A epoch by
10.5 days (see Fig.~\ref{composites}). It is clear that this delay 
produces a good match between the A and B light curves, especially in
the case of total flux density and polarization position angle at 15~GHz. 
An examination of the percentage polarization composite light curve
shows the match here to be poorer and suggests that the error bars in 
this case have been under-estimated. This is supported by the relatively
high $\overline{\chi}^2$ at best fit of the percentage polarization light
curves (see Table~\ref{delaytable}) and could explain the formal 1$\sigma$ 
discrepancy between the time delays derived from this data set and from the
polarization position angle data set.

To obtain an estimate of the Hubble constant, we model B0218+357 using 
the singular isothermal ellipsoid (SIE) mass model, as described in 
Kormann, Schneider \& Bartelmann (1994). As constraints on the mass model 
we use the 8.4~GHz flux density ratio A/B of 3.57 (Table~\ref{delaytable}), 
assuming an error of 0.015, which is approximately the 68.3 per cent 
confidence range. (Using the ratio 3.73 obtained from the 15~GHz data 
increases the predicted delay by $\sim$2 per cent.) As further constraints 
we use the 15~GHz VLBI observations of the mas structure of the images 
\cite{patnaik95}. Both components A and B consist of two subcomponents 
(A1,A2 and B1,B2) and all four are resolved. The flux density ratio is 
assumed to be the same for A1/B1 and A2/B2.  We also make use of the 
positions of the subcomponents as well as the deconvolved sizes and 
position angles of A1 and B1. Relative positional accuracies for the 
subcomponents of 0.1 mas are assumed. The centre of mass of the lens 
is a free parameter in the model.
 
\begin{figure}
\begin{center}
\setlength{\unitlength}{1cm}
\begin{picture}(5,6)(0,0)
\put(-0.5,-1.3){\includegraphics{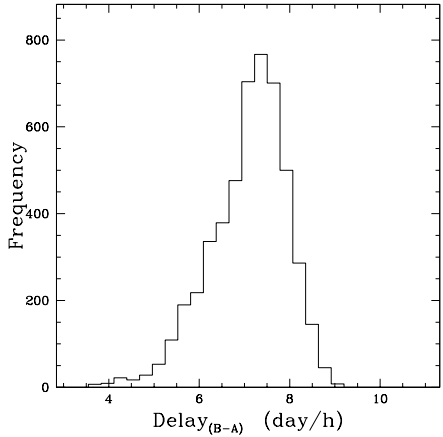}}
\end{picture}
\caption{Delay distribution function obtained for the lens model parameters
listed in Table~\ref{modeltab}.}
\label{hubblehist}
\end{center}
\end{figure}

We create 5000 `data sets', adding Gaussian-distributed errors to the 
observed positions, subcomponent sizes and flux density ratios. Using these 
artificial data sets, we solve for the minimum $\chi^2$ solution of the 
mass model parameters and the time delay between A1 and B1. The delay
error distribution is shown in Fig.~\ref{hubblehist} and the resulting 
median values of the mass model parameters are listed in 
Table~\ref{modeltab}. The errors indicate the range containing 68.3 per 
cent of the parameter distribution function.

\begin{table}
\begin{center}
\caption{Median values of mass model parameters with associated 68.3 per 
cent confidence limits. Lens centre coordinates are given relative to 
component A. Position angle is measured east from north.}
\begin{tabular}{cc}
Parameter & Value \\ \hline
Lens centre ($x$) & $252^{+15}_{-9}$ mas \\
Lens centre ($y$) & $115^{+4}_{-6}$ mas \\
Velocity dispersion & $170.7^{+2.0}_{-1.5}$ km\,s$^{-1}$ \\
Axis ratio & $0.77^{+0.08}_{-0.11}$ \\
Position angle & $-48^{\circ}$$^{+9}_{-19}$ \\
\end{tabular}
\label{modeltab}
\end{center}
\end{table}

The centre of mass obtained from the model coincides neither with the
centre of the radio ring nor with the galaxy centre derived from an
{\it HST} NICMOS observation (CASTLeS gravitational lens database,
http://pluto.harvard.edu/castles/). We do not regard these `offsets'
very seriously at present. An offset of the model centre from the
centroid of the radio ring emission could arise if, in the source
plane, the extended emission giving rise to the ring is asymmetrically 
placed with respect to the diamond caustic. An example of this is B1938+666 
\cite{king98}. Since existing radio maps do not define the ring structure 
very clearly it is difficult to tell if the ring is symmetric from the radio 
images. The galaxy centre obtained from the {\it HST} NICMOS image is also 
hard to interpret in the light of possible blending of the galaxy core with 
the B image \cite{xanthopoulos98}.

The model predicts a median delay of $7.2^{+1.3}_{-2.0}/h$ days
between A1 and B1; $H_0$ is equal to $100h$\,km\,s$^{-1}\,$Mpc$^{-1}$. 
The quoted errors indicate the range containing 95 per cent of the delay 
distribution function as shown in Fig.~\ref{hubblehist}. They are formal 
errors in the sense that they are based on the assumption that the effect 
of the lens is correctly modelled by the singular isothermal ellipsoid
potential, an assumption which will be testable in the future when further
modelling constraints are obtained from better radio maps of the
Einstein ring. Given this caveat, however, combining the model
results with the observed delay of $10.5\pm0.4$ days gives a value
for the Hubble constant of 69$^{+13}_{-19}$\,km\,s$^{-1}\,$Mpc$^{-1}$, where
the errors are the formal 95 per cent confidence limits. This is for 
$\Omega_{0} = 1$, $\lambda_{0} = 0$ and $\eta = 1$ (completely homogeneous 
universe). Since the source and lens redshifts are relatively low, even 
generous departures from these fiducial values have a negligible effect on 
the above value for $H_0$. 

The above value is consistent with those found from other lens systems, 
e.g. B0957+561 \cite{kundic97}, B1608+656 (Fassnacht et al., Koopmans \& 
Fassnacht, in preparation) and PKS 1830-211 \cite{lovell98} all of which 
lie within the range of approximately 60--70\,km\,s$^{-1}\,$Mpc$^{-1}$. 
The situation with the remaining lens system for which $H_0$ has been 
measured, PG 1115+080, is unclear as several models have been published 
with a large spread (42--80) in the corresponding values of $H_0$ 
\cite{schechter97,keeton97}. The value from the B0218+357 system is clearly 
capable of considerable improvement, mainly in refinement of the lens mass 
model, but also by increasing the accuracy of the delay determination.

\section*{Acknowledgements}

We thank Sunita Nair, Alok Patnaik and Patrick Leahy for useful discussions 
and Jaan Pelt for suggesting the simultaneous chi-squared minimization 
technique. ADB acknowledges the receipt of a PPARC studentship. We thank
Karl Menten for giving us 30 minutes of his VLA observing time. This research 
was supported in part by the European Commission TMR Programme, Research 
Network Contract ERBFMRXCT96-0034 `CERES' and has made use of data from the 
University of Michigan Radio Astronomy Observatory which is supported by the 
National Science Foundation and by funds from the University of Michigan. 
The VLA is operated by the National Radio Astronomy Observatory which is a 
facility of the National Science Foundation operated under cooperative 
agreement by Associated Universities, Inc.

\end{document}